\documentclass[%
 reprint,
superscriptaddress,
nofootinbib,
 amsmath,amssymb,
 aps,floatfix,
pra,
]{revtex4-2}

\usepackage{graphicx}
\usepackage{dcolumn}
\usepackage{bm}
\usepackage{cancel}
\usepackage{hyperref}

\usepackage[dvipsnames]{xcolor}
\usepackage[normalem]{ulem}
\begin{document}

\preprint{APS/123-QED}

\title{
A model for optimizing quantum key distribution with continuous-wave pumped entangled-photon sources
}

\author{Sebastian Philipp Neumann}
\email{sebastian.neumann@oeaw.ac.at}
 \affiliation{Institute for Quantum Optics and Quantum Information Vienna, Boltzmanngasse 3, 1090 Vienna, Austria}
 \affiliation{Vienna Center for Quantum Science and Technology, Boltzmanngasse 5, 1090 Vienna, Austria}

\author{Thomas Scheidl}
 \affiliation{Institute for Quantum Optics and Quantum Information Vienna, Boltzmanngasse 3, 1090 Vienna, Austria}
 \affiliation{Vienna Center for Quantum Science and Technology, Boltzmanngasse 5, 1090 Vienna, Austria}
\author{Mirela Selimovic}
 \affiliation{Institute for Quantum Optics and Quantum Information Vienna, Boltzmanngasse 3, 1090 Vienna, Austria}
 \affiliation{Vienna Center for Quantum Science and Technology, Boltzmanngasse 5, 1090 Vienna, Austria}
\author{Matej Pivoluska}
 \affiliation{Institute for Quantum Optics and Quantum Information Vienna, Boltzmanngasse 3, 1090 Vienna, Austria}
 \affiliation{Institute of Computer Science, Masaryk University, 602 00 Brno, Czech Republic}%
 \affiliation{Institute of Physics, Slovak Academy of Sciences, 845 11 Bratislava, Slovakia}%
 \author{Bo Liu}
 \affiliation{College of Advanced Interdisciplinary Studies, NUDT, Changsha 410073, P. R. China}
\author{Martin Bohmann}
 \affiliation{Institute for Quantum Optics and Quantum Information Vienna, Boltzmanngasse 3, 1090 Vienna, Austria}
 \affiliation{Vienna Center for Quantum Science and Technology, Boltzmanngasse 5, 1090 Vienna, Austria}
\author{Rupert Ursin}
\email{rupert.ursin@oeaw.ac.at}
 \affiliation{Institute for Quantum Optics and Quantum Information Vienna, Boltzmanngasse 3, 1090 Vienna, Austria}
 \affiliation{Vienna Center for Quantum Science and Technology, Boltzmanngasse 5, 1090 Vienna, Austria}

\date{\today}
\begin{abstract}
Quantum Key Distribution (QKD) allows unconditionally secure communication based on the laws of quantum mechanics rather than assumptions about computational hardness.
Optimizing the operation parameters of a given QKD implementation is indispensable in order to achieve high secure key rates.
So far, there exists no model that accurately describes entanglement-based QKD with continuous-wave pump lasers.
For the first time, we analyze the underlying mechanisms for QKD with temporally uniform pair-creation probabilities and develop a simple but accurate model to calculate optimal trade-offs for maximal secure key rates.
In particular, we find an optimization strategy of the source brightness for given losses and detection-time resolution.
All experimental parameters utilized by the model can be inferred directly in standard QKD implementations, and no additional assessment of device performance is required.
Comparison with experimental data shows the validity of our model.
Our results yield a tool to determine optimal operation parameters for already existing QKD systems, to plan a full QKD implementation from scratch, and to determine fundamental key rate and distance limits of given connections.
\end{abstract}
\maketitle
\section{\label{intro}Introduction}
    Quantum key distribution (QKD) is a method of creating a secret and random one-time pad for two remote users usable for unconditionally secure encryption of messages \cite{Gisin2002, Xu2020}.
    Since its first proposal in 1984 \cite{Bennett2014}, intense research has pushed QKD ever closer to real-life realizations.
    It has been shown via free-space links on ground \cite{Ursin2006, Yin2012, Ecker2021} and from space \cite{Yin2020} as well as for long-distance fiber links \cite{Wengerowsky2019} and in network configurations \cite{Wengerowsky2018a, Joshi2019}.
    Many different schemes have been proposed in recent decades, such as entanglement-based protocols (E91 \cite{Ekert1991} resp. BBM92 \cite{Bennett1992}), twin-field \cite{Chen2021arx} and decoy-state prepare-and-send implementations \cite{Lo05a}.
    Unlike prepare-and-measure protocols, entanglement-based applications have the advantage of being able to create their quantum states in a single coherent process based, for example, on spontaneous parametric down-conversion (SPDC). Therefore, no quantum random number generators or other electronical inputs are required. 
    Thus, provably no information about the individual photon state exists before the actual measurement.
    In this sense, entanglement-based protocols exploit the quantum nature of the correlations necessary for QKD on the most fundamental level and can be extended to device-independent QKD \cite{Acin2007}.
    QKD with entangled photons also allows quantum network configurations with many users using one and the same sending apparatus, an entangled-photon source (henceforth simply referred to as ``source") \cite{Joshi2019}.
    There are two fundamentally different ways to operate such a source: by creating the photon pairs with a continuous-wave (CW) or a pulsed pump laser.
    Up to now, no in-depth model exists for the prediction of key rates and the calculation of optimal source brightness for CW sources.
    A model describing sources pumped with a pulsed laser was published in 2007 \cite{Ma2007} and has been the state of the art ever since. In such pulsed schemes, all photon pairs are found in discrete and evenly spaced time modes depending on the laser's repetition rate.
    This rate can be tuned independently of the pulse intensity, allowing to individually address photon creation rate and multi-pair emission.
    Due to the broad frequency spectra in a pulsed-pump scheme, dispersion effects in the optics have to be accounted for, especially in the nonlinear crystals where the entangled photons are created.
    
    This model of pulsed operation can be applied to CW pumped sources with limited accuracy only, as will be shown below.
    CW pumping has several advantages compared to pulsed-pump schemes, especially in the context of fiber-based QKD: firstly, the spectrum of the down-converted photons is narrower, thus reducing dispersion effects in both source and transmission channels \cite{Neumann2021}. 
    Secondly, additional high-precision time synchronization is not needed as the temporal correlation peak can be precisely determined using a delay histogram.
    And thirdly, damage to the source optics due to high-intensity pulses can be avoided.
    
    In this work, we present for the first time a model that accurately describes CW-pumped entanglement-based QKD systems.
    Importantly, all necessary inputs to the model can be read directly from experimentally available data, without the need of any additional assumptions.
    Our approach allows to calculate optimal brightness values and coincidence window lengths as well as the resulting final key rate.
    Hence, the present results are of particular importance for state-of-the-art entanglement-based QKD applications.
    Comparison with experimental data demonstrates the validity of our model.
    Although we are focusing here on polarization-encoded BBM92 implementations, our approach can be extended to other degrees of freedom, which is, however, outside of the scope of this work.
    
    The paper is structured as follows: in Sec. \ref{sec:work}, we explain the basic working principle of polarization-encoded BBM92.
    We then develop our model in Sec. \ref{meat} by first introducing parameters for an idealized model (Sec. \ref{sec:ideal}), modifying them to account for experimental imperfections (Sec. \ref{sec:acc}) and then combining them into the final model to calculate the expected secure key rates (Sec. \ref{sec:err}).
    We optimize the key rate with regard to pair creation rate and temporal detection tolerance and compare our model with experimental data (Sec. \ref{sec:data}).
    Concluding, in Sec. \ref{sec:discuss} we discuss our findings and present optimal parameters to maximize key rates.    
    
\section{Working principle of entanglement-based QKD\label{sec:work}}
    Entanglement-based QKD protocols such as BBM92 \cite{Bennett1992} rely on entanglement between distant physical systems, in our case specifically in the polarization degree of freedom of a photon pair.
    In an idealized scenario, one can create maximally entangled photon pairs which form a so-called Bell state, e.g.,
    \begin{align}
        |\phi^+\rangle=\frac{1}{\sqrt{2}}(|H\rangle_\mathrm{A} \otimes |H\rangle_\mathrm{B}+|V\rangle_\mathrm{A} \otimes |V\rangle_\mathrm{B}).
        \label{eq:bellstate}
    \end{align}
    where $H$ ($V$) denotes horizontal (vertical) polarization and the subscripts signify the recipient of the single photon traditionally called Alice (A) and Bob (B).
    We choose this state because of the fact that it is correlated in the mutually unbiased linear polarization bases $HV$ and $DA$ (diagonal/antidiagonal), where $|D\rangle=\frac{1}{\sqrt{2}}(|H\rangle+|V\rangle)$ and $|A\rangle=\frac{1}{\sqrt{2}}(|H\rangle-|V\rangle)$.
    The following model can however be used for any Bell state, if the correlations are adapted accordingly.
    
    Alice and Bob measure their photons randomly and independently from each other either in the $HV$ or the $DA$ basis.
    The basis choice can in practice be realized actively or passively.
    Actively means that Alice and Bob switch their measurement bases depending on the outputs of a quantum random number generator.
    A QKD implementation with passive basis choice uses probabilistic beamsplitters to direct the photons to either a $HV$ or a $DA$ measurement, both of which are realized simultaneously.
    In the course of the paper, we will assume active basis choice unless noted otherwise.
    In any case, Alice and Bob record outcome ($H$, $D$=0 and $V$, $A$=1) and measurement basis for each event.
    By communicating about their measurement bases only, Alice and Bob can discard those recorded events where they measured in different bases and therefore see no correlation between their bit outcome (``sifting'').
    For the other events, they can expect perfect correlation, and thus use their sifted bit strings for key creation.
    By checking a randomly chosen subset of their sifted measurement outcomes to make sure that correlations have not degraded, Alice and Bob can rule out the existence of an eavesdropper.
    
    In a real experiment, however, perfect Bell states such as in Eq. \eqref{eq:bellstate} do not exist.
    The polarization correlations are degraded through optical imperfections of the source and the detectors, which result in bit and/or phase flips.
    Also, in practice it is not possible to distinguish each and every consecutively emitted entangled pair from one another due to imperfections in temporal detection, as discussed below.
    We call such temporally irresolvable emissions ``multipairs".\footnote{Unlike in pulsed-source BBM92 \cite{Ma2007}, coherent emission of $n$-pair states is negligible in the case of sources using CW pump lasers due to the photons' temporal multi-mode character \cite{Takesue2010}.}
    Multipairs degrade the quantum correlations necessary to create a secure key, since detection of a multipair photon at Alice does not unambiguously herald the detection of its entangled---and therefore perfectly correlated---partner photon at Bob (and vice versa).
    Instead, with a certain probability, the photon is wrongly identified as being correlated with a photon from another pair, which leads to errors.
    Based on these considerations, in what follows, we will define the parameters necessary to calculate the performance of a CW-QKD system.
    All of these parameters can easily be obtained from experimental detection results, thus making our model ideally suited for direct implementation in real-world applications.
    
\section{\label{meat}Modeling QKD with CW-pumped sources}
    
    For developing the model, we will start out with an idealized polarization-encoded CW-QKD protocol introducing the basic parameters (Sec. \ref{sec:ideal}).
    In Sec. \ref{sec:acc}, we will extend this consideration by taking into account noise counts and multipair effects.
    We then use the experimental quantities defined in this way to calculate error rate and secure key rate (Sec. \ref{sec:err}).

\subsection{Idealized CW-QKD system\label{sec:ideal}}
    The most general CW-pumped source setup uses a photon source creating an average number of entangled photon pairs per time unit.
    This quantity is called brightness $B$, for which we use the unit counts per second (cps) instead of Hertz to emphasize the random nature of the emission process.
    We assume the probability of photon-pair creation to be uniformly distributed in time, as is justified in the case of CW pumping \cite{Takesue2010}.
    
    The entangled photons are spatially separated and sent to communication partners Alice and Bob, where they are detected with overall channel probabilities $\eta_\mathrm{A}$ and $\eta_\mathrm{B}$, respectively.
    Although these probabilities are composed of the source's intrinsic heralding efficiency \cite{Klyshko1980}, the channel and coupling losses, the detection optics' transmission and the detectors' deadtimes and efficiencies, we will consider each $\eta_i$ as one single entity in the following calculations, sometimes referred to as system efficiency.
    This is because isolating individual loss effects is difficult in a real experiment and not required for our model.
    
    As a result of these definitions, the average local photon detection rate of Alice resp. Bob, the so-called single counts, can be written as
    \begin{align}
        S^\mathrm{t}_\textrm{A}=B\eta_\textrm{A} \textrm{ and } S^\mathrm{t}_\textrm{B}=B\eta_\textrm{B},
        \label{eq:St}
    \end{align}
    where we ignore noise counts for now.
    Note also that deadtime-induced losses, unlike other effects contributing to the $\eta_i$, are a function of detector count rates $S^\mathrm{t}_i$ and therefore of the brightness $B$, which has to be taken into account for low-loss scenarios (see Appendix \ref{sec:deadtime}).
    
    Naturally, two photons of a pair must be detected in order to observe their polarization correlation, i.e., use them for generating a cryptographic key.
    The rate of such two-photon events, which we call ``true coincident counts" or ``true coincidences" \footnote{Please note that \emph{true coincidences} are not a measurable quantity, since the experimenter cannot distinguish between a true or an accidental coincidence in principle. Even if an accidental coincidence does not conform to the expected correlations, it cannot be unambiguously identified as ``accidental", since ``true" coincidences can also be measured erroneously [cf. Eq. \eqref{eq:epol}]. Therefore, the notion of ``true" pairs is solely a useful concept for our model, describing the photons that actually provide the non-classical correlations necessary for QKD.}, is given as
    \begin{align}
        CC^\mathrm{t}=B\eta_\mathrm{A}\eta_\mathrm{B},
        \label{eq:CCt}
    \end{align}
    where we again preliminarily ignore noise counts.
    Using Eqs. \eqref{eq:St} and \eqref{eq:CCt}, the $\eta_i$ can be calculated as \cite{Klyshko1980}
    \begin{align}
        \eta_\mathrm{A}=\frac{CC^\mathrm{t}}{S^\mathrm{t}_\textrm{B}} \textrm{   and   }  \eta_\mathrm{B}=\frac{CC^\mathrm{t}}{S^\mathrm{t}_\textrm{A}}.
        \label{eq:heralding}
    \end{align}
    The $\eta_i$ are sometimes also called ``heralding efficiency'', since they give the probability that the detection of one photon in one arm announces, or ``heralds'', the detection of a photon in the other arm.
    One can also define a total heralding efficiency $\eta=\sqrt{\eta_\mathrm{A}\eta_\mathrm{B}}$.
    
    Imperfections of source, polarization compensation and optical detection system lead to erroneous polarization measurement outcomes, i.e. two-photon events which do not comply with the expected Bell state.
    We call the probability of such an erroneous measurement $e^\textrm{pol}$.
    It consists of contributions of the individual polarization error probabilities $e^\textrm{pol}_\textrm{A}$  and $e^\textrm{pol}_\textrm{B}$ of Alice and Bob, respectively:
    \begin{align}
        e^\textrm{pol}=e^\textrm{pol}_\textrm{A}(1-e^\textrm{pol}_\textrm{B})+e^\textrm{pol}_\textrm{B}(1-e^\textrm{pol}_\textrm{A}).
        \label{eq:epol}
    \end{align}
    It should be noted that measuring the wrong bit value at Alice \emph{and} Bob still counts as a valid measurement, since it is impossible in principle for the experimenter to distinguish such an event from a correctly measured true coincidence.
    In most practical implementations, it is more convenient to read $e^\textrm{pol}$ directly from the experimental data instead of quantifying the $e_i$ individually (see Appendix \ref{sec:para}).

\subsection{Noise-afflicted CW-QKD system \label{sec:acc}}
   In a real-world entanglement-based QKD implementation, the crucial source of error is not $e^\textrm{pol}$, which can be kept below 1\% in modern applications \cite{Anwar2021}, but the unavoidable registration of uncorrelated multipair photons which have lost their partner, and/or noise counts as coincidences.
   Such erroneous coincidences are called ``accidental coincidence counts".
    To calculate the accidental coincidence rate for BBM92 with a CW pump, firstly one needs to modify Eq. \eqref{eq:St} to account for dark counts $DC_i$ in the detectors:
    \begin{align}
       S^\mathrm{m}_\textrm{A}=S^\mathrm{t}_\textrm{A}+DC_\textrm{A} \textrm{ and } S^\mathrm{m}_\textrm{B}=S^\mathrm{t}_\textrm{B}+DC_\textrm{B}
       \label{eq:Sm}
    \end{align}
    where $S^\mathrm{m}_i$ are the actually measured count rates.
    Note that stray light, residual pump laser light, intrinsic detector dark counts or any other clicks which do not originate from source photons all have the same effect for our purposes. Therefore, we include all such clicks in the $DC_i$.
    In a real experiment, Alice and Bob require at least two detectors each to be capable of distinguishing orthogonal quantum states.
    In Eq. \eqref{eq:Sm}, we assume that Alice and Bob each own identical detectors whose photon and dark count rates can simply be added; for the case of non-identical detectors and polarization dependent detection efficiency, see Appendix \ref{sec:nonid}.
    
    Alice and Bob identify coincidences by looking for simultaneous detection times (accounting for a certain constant delay $t_\mathrm{D}$ caused by different photon travel times and electronical delays).
    There are three main effects that can degrade the fidelity of this identification: the detection system's finite timing precision, the coherence length of the photons, and chromatic dispersion effects in fiber, which delay photons of different wavelengths with respect to each other \cite{Neumann2021}.
    These effects cause a spread of the photons' temporal correlation function, whose full width at half maximum (FWHM) we call $t_\Delta$.
    Because in any real experiment $t_\Delta>0$, Alice and Bob need to define a so-called ``coincidence window'' $t_{CC}$. It can be understood as the temporal tolerance allowed for the difference in detection time of two correlated photons.
    
    It follows that there is a possibility of confusing uncorrelated detector clicks with true coincidences.
    This possibility can be calculated, since it depends on $t_{CC}$ and the $S^\mathrm{m}_i$.
    Assuming independent Poissonian photon statistics at Alice and Bob, one can define the mean number of clicks at Alice resp. Bob per coincidence window as
    \begin{align}
       \mu^S_\textrm{A}=S^\mathrm{m}_\textrm{A}t_{CC} \textrm{ and } \mu^S_\textrm{B}=S^\mathrm{m}_\textrm{B}t_{CC}.
       \label{eq:musi}
    \end{align}
    Most single-photon detectors used today are not photon-number resolving.
    Therefore, the chance of an accidental coincidence to be registered can be approximated by the probability of \emph{at least} one detection event taking place at each of them:
    \begin{align}
       P^\mathrm{acc}=(1-e^{-\mu^S_\textrm{A}})\cdot(1-e^{-\mu^S_\textrm{B}}),
       \label{eq:poissonPacc}
    \end{align}
    where we use the fact that the click probability is given by $(1-e^{-\mu^S_\textrm{i}})$; cf. \cite{Bohmann2017, Bohmann2019}.
    This expression for $P^\mathrm{acc}$ provides a good estimate for the accidental coincident-count probabilities in high-loss regimes.
    For low-loss scenarios it needs to be adapted as it overestimates the probability of accidental coincidence counts by also counting true coincidences as accidental (see Appendix \ref{sec:acclowloss}).
    For $\mu^S_i\ll1$, Eq. \eqref{eq:poissonPacc} can be simplified to
    \begin{align}
       P^\mathrm{acc}\approx\mu^S_\textrm{A}\cdot\mu^S_\textrm{B}. \label{eq:Pacc}
    \end{align}
    The rate of accidental coincidences per second is therefore
    \begin{align}
       CC^\mathrm{acc}=\frac{P^\mathrm{acc}}{t_{CC}} \approx\, \frac{\mu^S_\textrm{A}\cdot\mu^S_\textrm{B}}{t_{CC}}=S^\mathrm{m}_\textrm{A}\cdot S^\mathrm{m}_\textrm{B}\cdot t_{CC}.
       \label{eq:CCacc}
    \end{align}
    Note that since we assume \emph{at least} one detector click per receiver for an accidental count to happen, we take into account the fact that in a real experiment with several detectors, there can be more than one click per coincidence window (cf. \ref{sec:acclowloss}).
    In that case, a random bit value has to be assigned \cite{Luetkenhaus1999, Moroder2010}, which has the same error probability as an accidental count and can therefore be seen as a part of Eq. \eqref{eq:CCacc}.
    Also note that $CC^\mathrm{acc}$ depends quadratically on $B$, but $CC^\mathrm{t}$ linearly.
    Thus, noise increases faster than the desired signal when increasing $B$, which gives an intuitive understanding why simply pumping the source with higher power can only enhance the key rate up to a certain degree (see Sec. \ref{sec:data}).
    
\begin{figure}[b]
    \includegraphics[width=\columnwidth]{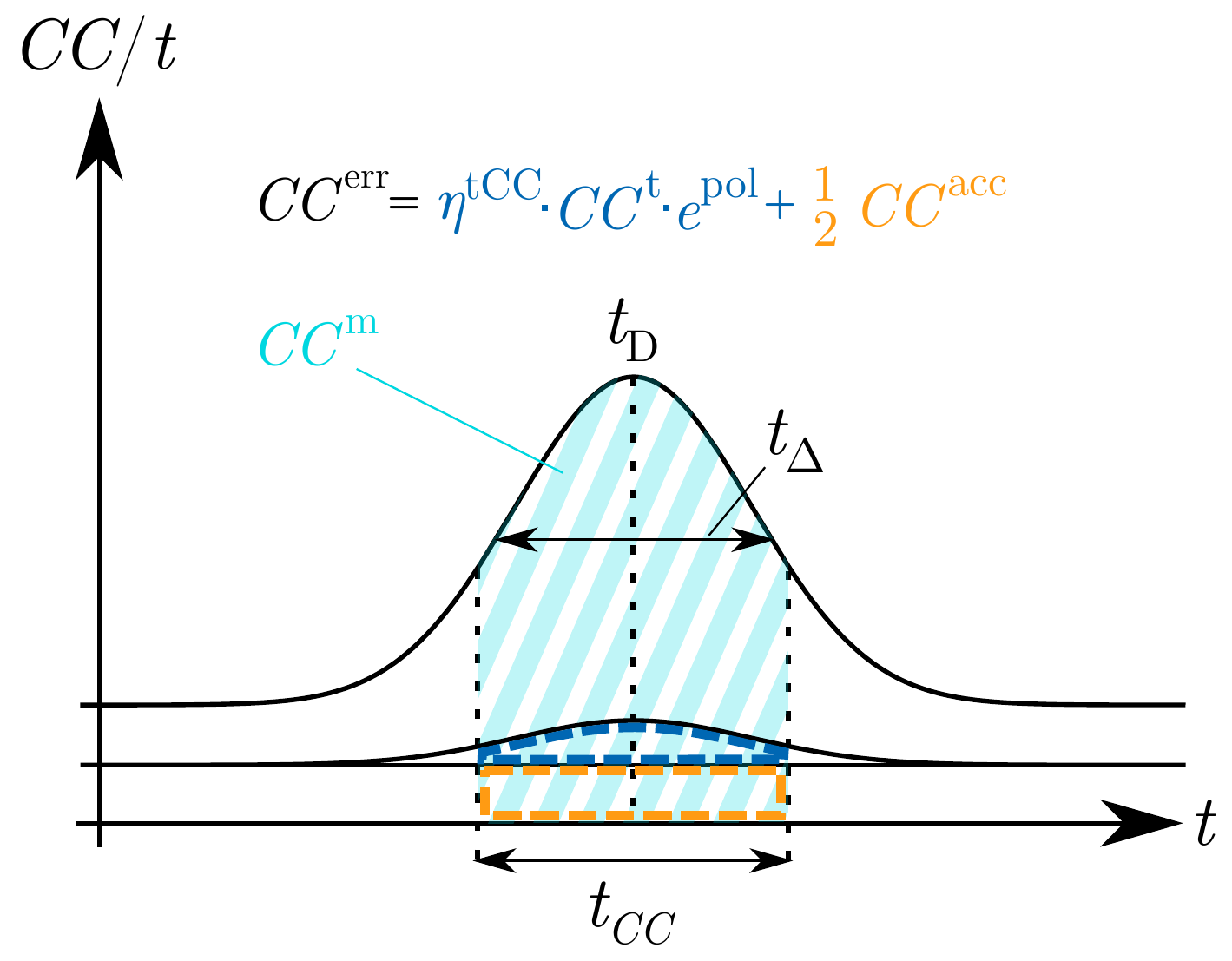}
    \caption{\label{fig:tCC}Number of coincidences per time unit for different relative measurement times. $t_\mathrm{D}$ is the delay between Alice and Bob and $t_\Delta$ is the FWHM of the temporal distribution, both of which are constant. The magnitude of the freely selectable coincidence window $t_{CC}$ not only determines the number of total coincidences $CC^\mathrm{m}$, but also the QBER $E$, i.e. the ratio of erroneous ($\eta^{t_{CC}}\cdot CC^\mathrm{t} \cdot e^\mathrm{pol}$) plus half of all accidental ($\frac{1}{2}CC^\mathrm{acc}$) coincidence counts to $CC^\mathrm{m}$.}
    \end{figure}

    It is not only accidental coincidences which depend on the choice of $t_{CC}$.
    If it is chosen in the order of the timing imprecision $t_\Delta$, true coincidences will be cut off and lost due to the Gaussian shape of the $g^{(2)}$ intensity correlation with FWHM $t_\Delta$ between Alice's and Bob's detectors (see Fig. \ref{fig:tCC}).
    
    This $g^{(2)}$ function can be modeled as a normal distribution
    \begin{align}
        j(t,t_\Delta,t_\mathrm{D})=\frac{2}{t_\Delta}\sqrt{\frac{\textrm{ln}(2)}{\pi}}\cdot \mathrm{exp}\Big[-\frac{4\,\mathrm{ln}(2)}{t_\Delta^2}\cdot(t-t_\mathrm{D})^2\Big]
    \end{align}
    with delay $t_\mathrm{D}$. 
    $t_\Delta$ is the resulting timing imprecision between Alice's and Bob's measurements, i.e., it is the convolution of detector jitter, chromatic dispersion and coherence time of the photons at both Alice and Bob.
    To arrive at the loss which true coincidences suffer due to the coincidence window, one can carry out the integration
    \begin{align}
\eta^{t_{CC}}&=\int\limits_{-t_{CC}/2}^{t_{CC}/2} j(t,t_\Delta,t_\mathrm{D}=0)dt=\\
&=\textrm{erf}\bigg[\sqrt{\textrm{ln}(2)}\cdot\frac{t_{CC}}{t_\Delta}\bigg].
\label{eq:etatcc}
\end{align}
    Here, $\eta^{t_{CC}}$ is the proportion of true coincidences which fall into the chosen coincidence window $t_{\mathrm{CC}}$ and are thus identified as coincidences in the experiment.
    In this sense, $\eta^{t_{CC}}$ can be interpreted as coincidence-window dependent detection efficiency.
    Now we can define the actually measured coincidences as
    \begin{align}
        CC^\mathrm{m}=\eta^{t_{CC}}CC^\mathrm{t}+CC^\mathrm{acc}.
        \label{eq:CCm}
    \end{align}
    This is the total number of detector events per second that Alice and Bob use to create their key.
    But obviously, a subset of these events occuring with rate $CC^\mathrm{err}$ actually does not show correlations in accordance with Eq. \eqref{eq:bellstate}: firstly, all those correlated photons which are measured erroneously; and secondly, on average half of all accidental coincidence counts:
    \begin{align}
        CC^\mathrm{err}=\eta^{t_{CC}}\cdot CC^\mathrm{t}\cdot  e^\mathrm{pol}+\frac{1}{2}CC^\mathrm{acc}.
        \label{eq:CCerr}
    \end{align}
    
\subsection{\label{sec:err}Error rate and secure key rate}
    
    From the quantities defined above, one can now calculate the quantum bit error rate (QBER $E$), i.e. the ratio of erroneous coincidences to total coincidences:
    \begin{align}
        E=\frac{CC^\mathrm{err}}{CC^\mathrm{m}}=\frac{\eta^{t_{CC}}\cdot CC^\mathrm{t}\cdot e^\mathrm{pol}+\frac{1}{2}CC^\mathrm{acc}}{\eta^{t_{CC}}\cdot CC^\mathrm{t}+CC^\mathrm{acc}}.
        \label{eq:QBER}
    \end{align}
 
    As a side remark, the commonly used parameter ``visibility'' $V$ relates to $E$ as $V=1-2E$ \cite{Gisin2002}.
    Fig. \ref{fig:tCC} shows a geometrical interpretation of Eq. \eqref{eq:QBER}.
    Coincidences correspond to different areas under the graphs, which are restricted by the chosen coincidence window.
    On one hand, it is desirable to increase the ratio of the light blue area to the combined dark blue and orange ones, which is equivalent to decreasing $E$.
    This can be done by decreasing $t_{CC}$, since the Gaussian-shaped $CC^\mathrm{m}$ (dark blue curve) scales more favorable in this case than the uniformly distributed accidental coincidence counts $CC^\mathrm{acc}$.
    On the other hand, reducing $t_{CC}$ means that $\eta^{t_{CC}}$ reduces the total number of coincidences which can be used for key creation.
    
    In order to evaluate the trade-off between these two effects, we will analyze the secret key rate in the limit of infinitely many rounds --- the so-called asymptotic key rate.\footnote{Where finite-key effects are of interest, one will have to take into account the total number of coincidences per block size and modify Eq. \eqref{eq:seckeyBBM92} accordingly. This might lead to different optimal experimental parameters satisfying Eq. \eqref{eq:dRs}. Nevertheless, the experimental parameter definitions of Sec. \ref{sec:acc} will be applicable also in this case.}
    Alice and Bob choose randomly between measurement settings in the $HV$ and $DA$ bases.
    Let us denote the probability that Alice and Bob measure in the same basis as $q$.
    Only in this case, the polarization measurement outcomes at Alice and Bob are correlated.
    All other coincidences have to be discarded.
    Therefore, the rate of coincidence rounds left for post-processing is equal to $q CC^m$.
    Subsequently, Alice and Bob reveal a small fraction of measurement outcomes in both bases to estimate the error.
    Now we can finally evaluate the amount of achievable key per second as \cite{Ma2007}:
        \begin{align}
          R^\mathrm{s} = qCC^m\Big[1-f(E_{\textrm{bit}})\mathrm{H}_2(E_{\textrm{bit}})-\mathrm{H}_2(E_{\textrm{ph}})\Big],
          \label{eq:seckeyBBM92}
        \end{align}
        where $\mathrm{H}_2$ is the binary entropy function defined as  
        \begin{align}
        \mathrm{H}_2(x)=-x\cdot\textrm{log}_2(x)-(1-x)\cdot\textrm{log}_2(1-x).\label{eq:h2}
        \end{align}
    $E_{\textrm{bit}}$ and $E_{\textrm{ph}}$ are the bit and phase error rates, which are measurement-basis-dependent rates of measurement outcomes incompatible with the maximally entangled state described in Eq. \eqref{eq:bellstate}. $f(E_{\textrm{bit}})$ is the bidirectional error correction efficiency which takes into account how much of the key has to be sacrificed due to the fact that post-processing is performed in finite blocks.
    In order to asses the validity of our model against an actual experiment, both the sifting rate $q$ and efficiency $f(E_{\textrm{bit}})$ need to be defined. We assume that the measurement settings of Alice and Bob are chosen uniformly, and thus $q =1/2$.
    Further, we choose a realistic value of $f(E_{\textrm{bit}})=1.1$ \cite{Elkouss2011}. Finally, since in our model the noise parameters are independent of measurement settings, we can set $E_{\textrm{bit}} = E_{\textrm{ph}} = E$. With these choices, key rate formula becomes:
    \begin{align}
        R^\mathrm{s}=\frac{1}{2}\cdot CC^\mathrm{m}\Big[1-2.1 \mathrm{H}_2(E)\Big].
        \label{eq:seckey}
    \end{align}
    From Eq. \eqref{eq:seckey} follows immediately that there is a fundamental limit $E_{max}\approx0.102$, above which no key creation is possible.
    In the following section we maximize $R^\mathrm{s}$ depending on the parameters discussed up to now.
    Importantly, all parameters used in this optimization can be directly determined in real-life experiments, which is explained in detail in Appendix \ref{sec:para}.
    Finally, note that the key rate formula can be adjusted using Eq. \eqref{eq:seckeyBBM92} to take into account measurement setting dependent losses as well; cf. Appendix \ref{sec:rate} for details.

\section{Comparison to experimental data \label{sec:data}}
    
\begin{figure*}[htbp!]
    \includegraphics[width=\textwidth]{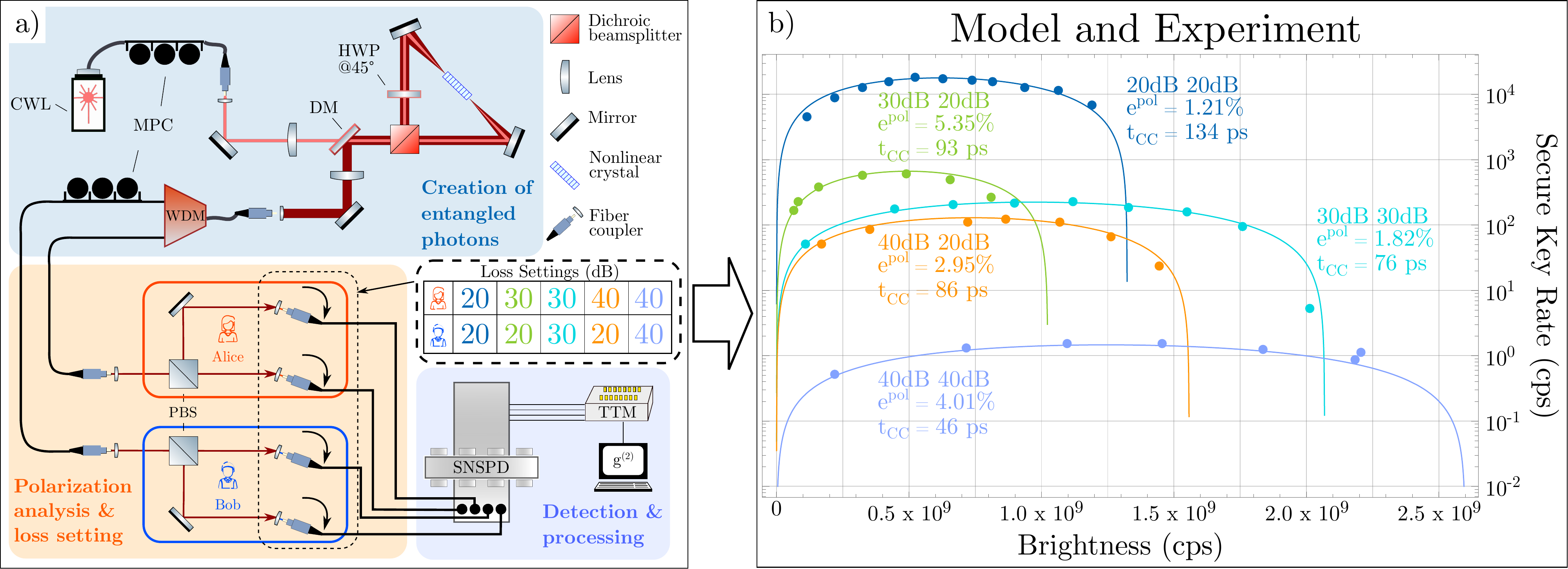}
    \caption{\label{fig:experiment} a) Setup used to create and detect polarization-entangled photon pairs for quantum key distribution. Top: A periodically poled lithium niobate nonlinear crystal is placed inside a Sagnac-type interferometer loop and pumped bidirectionally with a 775 nm continuous-wave laser (CWL). Via spontaneous parametric down-conversion (SPDC), it produces $H$-polarized photon pairs. The polarization of the counterclockwise pair is rotated to $V$ via a half-wave plate (HWP) set to 45°. It interferes with the clockwise pair at the beamsplitter and is directed to a single-mode-fiber (SMF) coupler by use of a dichroic mirror (DM). An off-the-shelf wavelength division demultiplexer (WDM) separates the two photons and directs them to two polarization analysing modules (bottom left). The manual polarization controller (MPC) in the pump fiber is used to set the photon pairs' Bell state of Eq. \eqref{eq:bellstate}. The MPC in Alice's arm compensates for the random polarization rotation in Bob's arm in order to arrive at the desired correlations. Alice and Bob perform an orthogonal polarization state measurement on their photon using polarizing beamsplitters (PBS) whose output modes are coupled into SMF and directed to superconducting nanowire single-photon detectors (SNSPD). Different loss scenarios are set by purposeful misalignment of the fiber couplers. Detection events of the SNSPD channels are recorded using a time-tagging module (TTM). From these tags, the $g^{(2)}$ correlation function can be determined from delay histograms between the channels, and secure key rates can be calculated. b) Comparison of our model (solid lines) and experimentally obtained data points (dots) for different loss settings and polarization measurement errors $e^\mathrm{pol}$. SNSPD jitter values vary with count rate, which we account for in the model calculations by making $t_\Delta$ a linear function of $B$. The data show that our model correctly predicts secure key rates over a wide range of losses, polarization errors and brightness values.}
    \end{figure*}

    For realistic applications, the $\eta_i$, the optical error $e^\mathrm{pol}$, the dark counts $DC_i$ and the temporal imprecision $t_\Delta$ cannot be modified freely.
    Two important parameters however can be chosen by the experimenter: brightness $B$ and coincidence window $t_{CC}$.
    The experimenter can vary $B$ up to a certain level by changing the laser pump power in the source.
    With laser powers of many hundreds of milliwatts, brightness values of up to $10^{10}$\,cps are feasible with current state-of-the-art sources \cite{Anwar2021}.
    The coincidence window $t_{CC}$ can in principle be chosen at will. 
    It follows that for each QKD scenario, there is an optimal choice of $B$ and $t_{CC}$ which maximizes $R^\mathrm{s}$ of Eq. \ref{eq:seckey}. Fig. \ref{fig:experiment} shows a comparison of our model and experimental values, where $t_{CC}$ has been numerically optimized for each curve with regard to the highest obtainable key rate and is then kept constant for every curve.

    The data were collected using a Sagnac-type source of polarization entangled photons in the telecom C-band.
    For a detailed description of such a source's working principle, we refer the reader to Ref. \cite{Anwar2021}.
    After passing wavelength division multiplexing (WDM) filters of 18.4\,nm FWHM centered about 1531 and 1571\,nm, the photons  impinge on single-photon superconducting nanowire detectors (SNSPDs) of the Single Quantum Eos Series with detection efficiencies of 80\% and deadtimes as low as 40\,ns according to the manufacturer.
    The detectors were connected to a time tagging module (TTM) Ultra 8 by Swabian Instruments.
    To keep the analysis of the model simple, we measured only in one superposition basis.
    Losses were introduced by controlled misalignment of the single-mode-fiber (SMF) couplers.
    All experimental parameters were determined by using count rates, coincidence rates and temporal histograms of the single-photon detections only, with no need of additional ``external" characterization (cf. Appendix \ref{sec:para}).
    Since the timing jitter of nanowire detectors strongly depends on the count rates they measure, linear fits of the jitter change depending on brightness have been included in the model.
    
    The data show excellent agreement with our model's predictions.
    The losses introduced in the measurements range from $40$ to $80$\,dB in total, with different distributions along the channels.
    Note that the two loss scenarios with equal total loss of $60$\,dB (orange and turquoise curve) perform very differently.
    Assuming $DC_A=DC_B$ \footnote{If $DC_A$ and $DC_B$ differ strongly, loss asymmetry can actually be beneficial. In a simplified view, this is because higher dark count rates matter less when occurring at detectors with higher single count rates. However, since in most scenarios neither loss nor dark counts can be chosen freely, we omit an in-depth discussion of this effect.}, symmetric loss is preferable to asymmetric loss because the probability of a partnerless photon matching with a dark count is reduced in this case.
    In Fig. \ref{fig:experiment}, this effect on the two $60$\,dB curves is, however, exaggerated due to different polarization errors $e^\mathrm{pol}$, which we set via a manual polarization controller (MPC) to show the model's validity for different parameter regions.
    The total losses are equivalent to in-fiber distances between 200 and 400\,km.
    Nevertheless, our model can be applied to all kinds of quantum channels, including e.g. free-space satellite connections, where variation of the channel attenuation \cite{Vasylyev2016, Bohmann2017a} can be integrated in our model in a straightforward manner.

    We want to emphasize that in any case, our optimization strategy works exclusively with experimentally measurable quantities that can be inferred directly from the actual QKD implementation (see Appendix \ref{sec:para}).
    Furthermore, the presented model can be used during the planning phase of an experiment to devise optimal working parameters based on specification sheets.
    While several calculations are approximated in our model, it shows excellent agreement with the experimental data.
    This is proof of its usefulness in a wide range of experimental parameters.
    For a more extensive treatment of phenomena that might become necessary in certain parameter regimes, such as dead time effects, low-loss channels and non-identical detectors, we refer the reader to Appendix\,\ref{sec:addcorr}.
    
\section{Optimization of QKD with a CW-pumped source \label{sec:discuss}}
 \begin{figure}[htbp!]
    \includegraphics[width=\columnwidth]{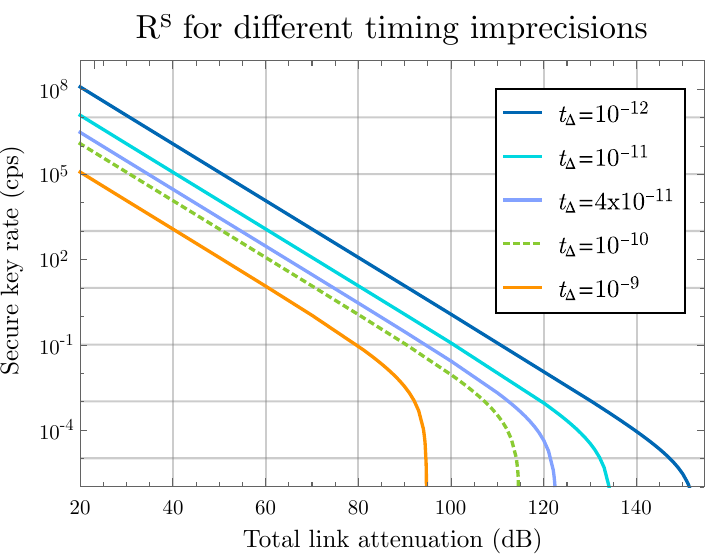}
    \caption{\label{fig:Rsvslossjit}Key rate $R^\mathrm{s}$ vs. total symmetric link loss $\eta_\mathrm{A}\cdot\eta_\mathrm{B}$ for different timing imprecision values $t_\Delta$. Brightness $B$ and coincidence window $t_{CC}$ have been optimized for every point of every curve. Dark counts $DC_\mathrm{A}=DC_\mathrm{B}=250$\,cps are kept constant for each of the four detectors per communication partner. Also polarization error $e^\mathrm{pol}=1\%$ is constant for all curves. Lower $t_\Delta$ allows both for higher key rates and longer maximum distance, since $CC^\mathrm{acc}$, the main source of errors, is directly proportional to $t_\Delta$. Note that the dotted green curve ($t_\Delta=10^{-10}$\,ps) is the same curve as the equally colored one in Fig. \ref{fig:RsvslossDC}.}
\end{figure}
\begin{figure}[htbp!]
    \includegraphics[width=\columnwidth]{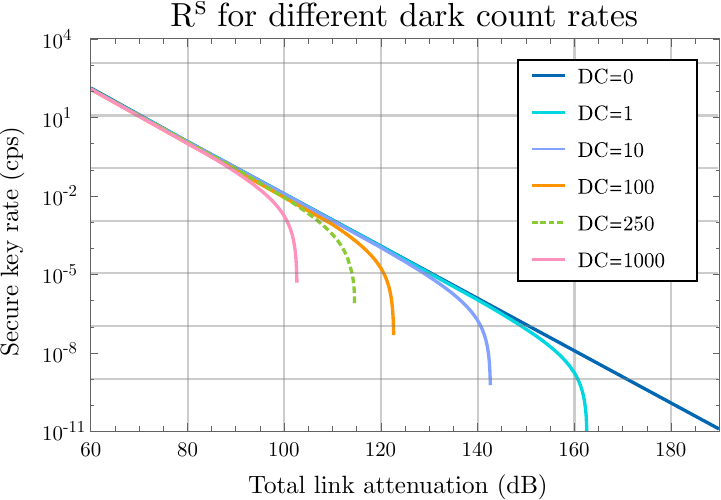}
    \caption{\label{fig:RsvslossDC} Key rate $R^\mathrm{s}$ vs. total symmetric link loss $\eta_\mathrm{A}\cdot\eta_\mathrm{B}$  for different dark count rates $DC$ per detector and four detectors per communication partner. The timing imprecision $t_\Delta$ is kept constant at $100$\,ps and the polarization error at $e^\mathrm{pol}=1\%$. As can clearly be seen, reducing detector noise counts effectively only increases the maximum achievable distance. In the case of no dark counts (dark blue curve), there exists no distance limit, since $t_{CC}$ can in principle be set arbitrarily small, thus keeping the error rate below $E_\mathrm{max}$ for any loss.
    Note that the dotted green curve ($DC=250$) is the same curve as the equally colored one in Fig. \ref{fig:Rsvslossjit}.}
    \end{figure}

    Now that we have shown the validity of our model in different parameter scenarios, we want to use it to illustrate limits and potential of CW-QKD.
    Therefore, we numerically maximize both $B$ and $t_{CC}$ for every point on the curves in Figs. \ref{fig:Rsvslossjit} and \ref{fig:RsvslossDC}, i.e.,
    \begin{align}
        \frac{\partial}{\partial B}\frac{\partial}{\partial t_{CC}}R^\mathrm{s}(B, t_{CC}; \eta_i, e^\mathrm{pol}, DC_i, t_\Delta)=0
        \label{eq:dRs}
    \end{align}
    is fulfilled continuously.
    Fig. \ref{fig:Rsvslossjit} shows the maximum obtainable key rate assuming symmetric loss for different jitter values.
    Lower jitter allows for a smaller coincidence window, which in turn allows for higher brightness values and thus key rates.
    Note that no matter the jitter value, there is an abrupt drop to zero key after a certain amount of loss.
    This is because dark counts will inevitably induce a minimum accidental coincidence count value $CC^\mathrm{acc}_\mathrm{min}=DC_\mathrm{A}\cdot DC_\mathrm{B}\cdot t_{CC}$.
    In a regime of high loss, this constant value can mask true coincidences if $\eta^{t_{CC}}\cdot CC^\mathrm{t}\lesssim10\cdot CC^\mathrm{acc}_\mathrm{min}$.
    In this case, key creation is frustrated.
    Fig. \ref{fig:RsvslossDC} now shows how $CC^\mathrm{acc}_\mathrm{min}$ is reduced with lower dark count values.
    For the hypothetical case of $DC_i=0$, the accidental coincidences $CC^\mathrm{acc}$ can be decreased to arbitrarily low values by reducing the brightness $B$. Although this also decreases maximum key rates beyond the point of usefulness, they never drop to zero, as indicated by the dark blue curve.
    When comparing Fig. \ref{fig:Rsvslossjit} and \ref{fig:RsvslossDC}, it becomes apparent that in a real-world scenario, reducing the timing imprecision $t_\Delta$ is more important than reducing the dark counts.
    This is because lower $DC_i$ can only increase the maximum distance in high-loss regimes, where key rates are extremely low already.
    To increase the key rate for a given loss, it is more favorable to lower $t_\Delta$ in most cases.
    
\begin{figure}
    \includegraphics[width=\columnwidth]{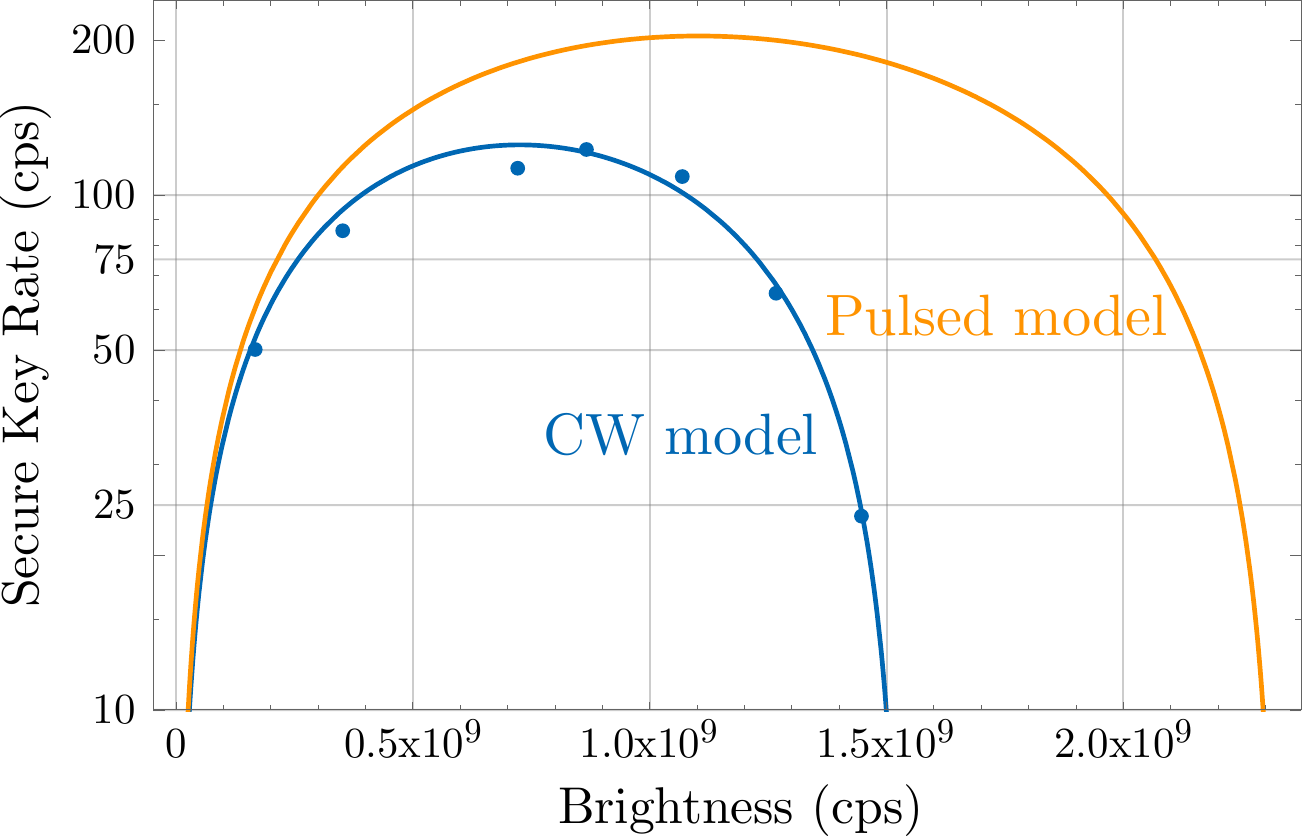}
    \caption{\label{fig:MavsUs} Comparison of 40dB 20dB experimental data with the secure key predicted by our model (blue line) versus an adapted version of the pulsed-source model from \cite{Ma2007} (orange line). Since temporal detection imprecision does not enter the pulsed-source model, it overestimates both the maximum key rate and the optimal brightness value.}
    \end{figure}

    We would also like to emphasize that when wrongly using the model for pulsed-source BBM92 by Ma et al. \cite{Ma2007} to estimate key rates for a CW-pumped implementation, one arrives at erroneous results, even when trying to adapt it.
    One could try to do so by replacing the mean photon number per pulse $2\lambda$ with the average photon number per coincidence window $\mu=B\cdot t_{CC}$ and changing the multipair probability of Eq. (5) in Ref. \cite{Ma2007} to a Poissonian distribution.
    Since doing so ignores any effects of temporal uncertainty, the results differ strongly, as can be seen in Fig. \ref{fig:MavsUs}.
\section{Conclusion}
    To the best of our knowledge, we have for the first time presented a comprehensive and accurate model of continuous-wave entanglement-based quantum key distribution.
    Our model allows to estimate and optimize the performance of any given CW-QKD system by extracting experimental parameters from the recorded detections only, without the need to perform any additional characterization of the experiment.
    It also allows to compare different devices and find the optimal solution for a given quantum link.
    For a given QKD setup, the model can accurately estimate the optimal settings of brightness and coincidence window to extract the maximal possible key and thus enhance the performance of the implementation.
    Furthermore, the presented approach is readily extendable to BBM92 based on entanglement in other degrees of freedom.
    We are confident that our easy-to-implement model will be used as an important design and optimization tool for CW-QKD links.
    
\begin{acknowledgments}
    We acknowledge European Union’s Horizon 2020 programme grant agreement No. 857156 (OpenQKD) and the Austrian Academy of Sciences. M.P. additionally acknowledges the support of VEGA project 2/0136/19 and GAMU project MUNI/G/1596/2019. B.L. acknowledges support of the National Natural Science Foundation of China under Grant No. 61972410 and the Research Plan of National University of Defense Technology under Grant No. ZK19-13.
\end{acknowledgments}

\appendix

\section{Parameter estimation \label{sec:para}}

There are numerous ways to estimate the parameters discussed in this work.
When planning a QKD link from scratch, one has to rely on data sheets and fiber loss measurements.
However, one can also estimate all parameters with the same QKD equipment used for the experiment, if already available.

\begin{figure}[htbp!]
    \includegraphics[width=\columnwidth]{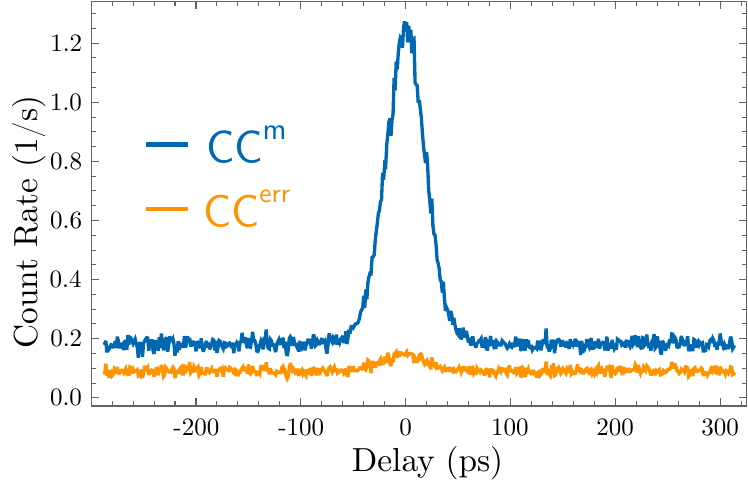}
    \caption{\label{fig:realpeak} Measured $g^{(2)}$ correlation histogram for the 40dB 40dB loss setting. $CC^\mathrm{m}$ contains all measured counts for all possible channel combinations, where all histograms have been shifted by their respective delays $t_D$. $CC^\mathrm{err}$ only shows undesired correlations, i.e. between polarization measurements not in accordance with Eq. \eqref{eq:bellstate}. The orange curve's small peak around 0 corresponds to erroneous polarization measurements, while the noise floor is equivalent to accidental coincidence counts $CC^\mathrm{acc}$ (cf. Fig. \ref{fig:tCC}).}
\end{figure}

Directly accessible parameters for the experimenter are $t_{CC}$ (since it is a free variable to be chosen by the experimenter), the $S^\mathrm{m}_i$ and $CC^\mathrm{m}$.
The \textbf{delay $t_\mathrm{D}$} between Alice's and Bob's detection times can be inquired by calculating a delay histogram of single counts at Alice and Bob and determining the location of the histogram peak (see Fig. \ref{fig:realpeak}).
From the same histogram, the \textbf{(total) timing imprecision $t_\Delta$} can be read from the peak's FWHM (less $CC^\mathrm{acc}$).
It should be mentioned that SNSPD jitter depends on both the detector's bias current and its count rate, and exhibits the lowest specified values for high current and low count rates only.
This dependency has been included in the model of Fig. \ref{fig:experiment} by using a linear fit of $t_\Delta$ vs. $B$ rather than a constant jitter value.

The \textbf{dark counts} $DC_i$ can be determined by blocking the source of photons and observing the $S^\mathrm{m}_i$, which are equal to the $DC_i$ for $B=0$ [see Eqs. \eqref{eq:St} and \eqref{eq:Sm}].
Note however that stray light from the pump beam cannot be observed with this method.
To do so, one either needs filters that block just the SPDC wavelength, or the possibility to frustrate SPDC without blocking or misdirecting the laser, e.g. by changing the crystal temperature.
Especially for long-distance single-mode-fiber links designed for the SPDC wavelength, it is safe to assume that pump light is sufficiently suppressed at the detectors.

For the following calculations, it is necessary to determine $CC^\mathrm{t}$ (for a certain brightness).
Especially in the case of low loss and low jitter, this can be done experimentally by lowering the brightness to a value where $CC^\mathrm{acc}\rightarrow0$ and therefore $CC^\mathrm{m}\rightarrow CC^\mathrm{t}$.
Alternatively, $CC^\mathrm{acc}$ can be subtracted from $CC^\mathrm{m}$: either by calculation using Eq. \eqref{eq:CCacc} or experimentally by changing $t_D$ to a value far from the actual coincidence peak, while keeping $t_{CC}$ constant.
In absence of $CC^\mathrm{t}$, the measured $CC^\mathrm{m}$ become equal to $CC^\mathrm{acc}$.
For all these approaches, it is important to choose $t_{CC}$ large enough such that $\eta^{t_{CC}}\rightarrow1$; as a rule of thumb, $t_{CC}=3\cdot t_\Delta$ is sufficient.

Now to determine the \textbf{optical error} $e^\mathrm{pol}$, one can use the methods just described to eliminate $CC^\mathrm{acc}$ in Eq. \eqref{eq:QBER} such that $E\approx e^\mathrm{pol}$.

The \textbf{heralding efficiencies} or transmission factors $\eta_i$ can be calculated using Eq. \eqref{eq:heralding}, where again, $CC^\mathrm{t}$ and $S^\mathrm{t}_i$ have to be determined in advance by subtracting $CC^\mathrm{acc}$ and $DC_i$.

Finally, also the \textbf{brightness} $B$ can be calculated using $CC^\mathrm{t}$ and $S^\mathrm{t}_i$ via
\begin{equation}
    B=\frac{S^\mathrm{t}_\mathrm{A}\cdot S^\mathrm{t}_\mathrm{B}}{CC^\mathrm{t}}.
\end{equation}
Note that for this calculation of the $\eta_i$ and $B$, deadtime effects have not been taken into account. Thus, even if the $CC^\mathrm{acc}$ are simply measured and subtracted, one should take care to operate the source at sufficiently low pump power (see Appendix \ref{sec:deadtime}).

If it should be necessary to incorporate \textbf{deadtime effects}, the most efficient way to determine $t_\dagger$ is to calculate an auto-correlation histogram in time of each detector channel while subjecting it to photons with Poissonian emission statistics. The temporal stretch for which no correlations are found is the detector channel's deadtime.

\section{Additional corrections \label{sec:addcorr}}

\subsection{Deadtime loss \label{sec:deadtime}}
In scenarios with high detector count rates, an additional loss factor might be considered to account for the detectors' deadtime $t_\dagger$ \cite{Becares2012}:

\begin{align}
    \eta^{t_\dagger}_i=\frac{1}{1+B\eta_it_\dagger/d}.
\end{align}
Here $d$ is the number of (identical, cf. Appendix \ref{sec:nonid}) detectors deployed per communication partner.
This effective loss cannot simply be considered as a constant contribution to $\eta_i$, since it is a function of $S^m_i$ and therefore $B$.
For $B\eta_i\cdot t_\dagger/d<0.02$, $\eta^T_i\approx1$ holds. Note that the estimation of $B$ can be compromised if this assumption is not justified due to low loss, high brightness and/or long detector deadtime.

Another result of deadtime loss is that the definition of the $\mu^S_i$ in Eq. \eqref{eq:musi} needs to be modified, since photons arriving at the detectors during the deadtime do not contribute to $S^\mathrm{m}_i$. One therefore needs to modify the $CC^\mathrm{acc}$ in Eq. \eqref{eq:CCacc} to
\begin{align}
    CC^\mathrm{acc}_{t_\dagger}\approx\frac{S^\mathrm{m}_\mathrm{A}\cdot S^\mathrm{m}_\mathrm{B} \cdot t_{CC}}{\eta^{t_\dagger}_\mathrm{A}\cdot \eta^{t_\dagger}_\mathrm{B}}
\end{align}
where we assume $DC_i\ll S^\mathrm{t}_i\eta_i$, which is reasonable in the high single-count regimes where deadtime effects become important.

\subsection{Accidental coincidence probability \label{sec:acclowloss}}
Equation \eqref{eq:poissonPacc} slightly overestimates the probability of accidental coincidence counts.
Since it assumes completely independent photon statistics at Alice and Bob, \emph{any} photon contributes to $CC^\mathrm{acc}$, regardless of whether it has lost its partner or not.
Thus, here we want to give a more extensive description $P^{\mathrm{acc}}_\mathrm{ext}$, which is well approximated by $P^{\mathrm{acc}}$ in Eq. \eqref{eq:Pacc} for $\eta_i\ll1$.
We start by defining the probability of a coincidence happening per coincidence window, $P^{CC^t}$:
\begin{widetext}
\begin{flalign}
P^{CC^t}=
\sum^\infty_{n=1}
    e^{-\mu}\frac{\mu^n}{n!}
    \sum^n_{i=1}
        \bigg[
            \Big(&
        (1-\eta_\mathrm{A})^{i-1}(1-\eta_\mathrm{B})^{i-1}\eta_\mathrm{A}\eta_\mathrm{B}
            \Big)\times\nonumber\\
            \times\Big(&1-\frac{\eta_\mathrm{A}}{2}
            \Big)^{n-i}
            \Big(1-\frac{\eta_\mathrm{B}}{2}
            \Big)^{n-i}\times\nonumber\\
            \times\Big(&
            1  -
                \frac{P^{DC}_\mathrm{A}}{2}\Big) \Big(1-\frac{P^{DC}_\mathrm{B}}{2}
            \Big)
        \bigg]
        \label{eq:PCCt}
\end{flalign}
\end{widetext}
where $\mu=B\cdot t_{CC}$ is the average number of photon pairs created per coincidence window before any loss, and $P^{DC}_i=DC_i\cdot t_{CC}$ are the probabilities of a noise count happening at Alice resp. Bob per coincidence window.
This formula takes into account the Poissonian emission and dark count statistics.
Multi-pair emissions can still yield a valid measurement if photons get lost in a way that two correlated photons end up at the detectors before all others (first factor inside the square brackets).
However, if photons emitted after the true pair, but inside the coincidence window, are detected as well, they can in some cases eliminate a true coincidence (second line). The divisions by $2$ come from the fact that if the later photon detection would occur in the same detector as the true photon detections, this case cannot be distinguished from a true coincidence.
If it clicks in the other detector, a random bit value has to be assigned, i.e. only this case has to be counted as an accidental.
Dark counts can also occur in the presence of a true pair, eliminating a valid coincidence in the same way as photons arriving later, which gives rise to the factors in the third line.
As a side remark, in the case of passive basis choice using beamsplitters, there are 4 instead of 2 detectors deployed; accordingly, the factor $1/2$ has to be replaced by $3/4$.

Using $P^{CC^t}$, the actual probability of detecting an accidental coincidence per coincidence window reads
\begin{widetext}
\begin{flalign}
P^\mathrm{acc}_\mathrm{cor}=
1-&
e^{-\mu}\Big[1-P^{DC}_\mathrm{A}\cdot P^{DC}_\mathrm{B}\Big]-P^{CC^t}-
\nonumber\\
-&\sum^\infty_{n=1}
    e^{-\mu}\frac{\mu^n}{n!}\Bigg[
        (1-\eta_\mathrm{A})^n+(1-\eta_\mathrm{B})^n-(1-\eta_\mathrm{A})^n(1-\eta_\mathrm{B})^n-\nonumber\\
        -&(1-\eta_\mathrm{A})^{n}(1-(1-\eta_\mathrm{B})^{n})\cdot P^{DC}_\mathrm{A}-(1-(1-\eta_\mathrm{A})^{n})(1-\eta_\mathrm{B})^{n}\cdot P^{DC}_\mathrm{B}-\nonumber\\
        -&(1-\eta_\mathrm{A})^{n}(1-\eta_\mathrm{B})^n\cdot P^{DC}_\mathrm{A}\cdot P^{DC}_\mathrm{B})
        \Bigg]
        \label{eq:CCtext}
\end{flalign}
\end{widetext}
The formula can be understood as follows: The accidental coincidence probability $P^\mathrm{acc}_\mathrm{cor}$ can be seen as all those two-click events that did not originate from a true pair.
We proceed by subtracting from probability $1$ all events which are no accidental coincidences.

Thus, in the first line, we subtract the probability of no photon pair being emitted, corrected by the case of two dark counts producing a coincidence. We also subtract all correct coincidences according to Eq. \eqref{eq:PCCt}.
Then we subtract the sum over all remaining pair emission probabilities which are not the vacuum state, not a true coincidence and not an accidental count. In the second line, we count those cases where no accidental coincidence happens since in at least one arm, no click occurs.
Since the possibility of both detectors not clicking is included in both $(1-\eta_A)^n$ and $(1-\eta_B)^n$, it has to be subtracted.
This subtraction avoids mistakenly counting the case of all photons lost twice.

In lines three and four of Eq. \eqref{eq:CCtext}, we have to re-add the cases where dark counts cause an accidental coincidence by ``replacing" a photon.
All other dark count cases are already included in the first line of the equation---either as part of $P^{CC^t}$ or in $1$, since a dark count happening when an accidental coincidence would have occurred anyway does not change their statistics.

For $\eta_i\ll1$, one can approximate $P^\mathrm{acc}_\mathrm{cor}$ with $P^\mathrm{acc}$ from Eq. \eqref{eq:Pacc}, which actually constitutes an upper bound for Eq. \eqref{eq:poissonPacc}.

\subsection{Non-identical detectors \label{sec:nonid}}

In our model, we assume Alice resp. Bob to use identical detectors for their orthogonal polarization measurements.
It has recently been shown \cite{Zhang2021} that vast differences in detector performance do not necessarily degrade the security of a QKD protocol.
However, different detection efficiencies lead to asymmetric single-count rates and therefore different accidental coincidence rates for different polarization correlations.
On top of this, different detector jitters lead to different $\eta^{t_{CC}}$ for each correlation.
These asymmetries and differences of used detectors can lead to a deviation from the reported model. 

To account for such imbalances one has to define two heralding efficiencies per communication partner, which we denote by $\eta_{\mathrm{A}j}$ and $\eta_{\mathrm{B}k}$, where $j$ and $k$ indicate the detectors.
Following Eq. \eqref{eq:CCt}, one can now differentiate true coincidence values:
\begin{align}
   CC^\mathrm{t}_{jk}=B\eta_{\mathrm{A}j}\eta_{\mathrm{B}k},
\end{align}
for which
\begin{align}
    \sum_{j,k=1}^{2}CC^\mathrm{t}_{jk}=CC^\mathrm{t}
\end{align}
holds. 
Additionally, one has to subdivide the $S^\mathrm{m}_i$ while accounting for different dark count rates
\begin{align}
   S^\mathrm{m}_{\mathrm{A}j}=B\eta_{\mathrm{A}j}+DC_{\mathrm{A}j}\\ S^\mathrm{m}_{\mathrm{B}k}=B\eta_{\mathrm{B}k}+DC_{\mathrm{B}k}
\end{align}
where similarly
\begin{align}
    S^\mathrm{m}_\mathrm{A}=\sum^{2}_{j=1} S^\mathrm{m}_{\mathrm{A}j} \mathrm{\hspace{0.2cm}and\hspace{0.2cm}} S^\mathrm{m}_\mathrm{B}=\sum^{2}_{k=1} S^\mathrm{m}_{\mathrm{B}k}
\end{align}
and assign different accidental coincidence rates to different detector combinations:
\begin{align}
    CC^\mathrm{acc}=\sum^{2}_{j,k=1} CC^\mathrm{acc}_{jk}=\sum^{2}_{j,k=1} S^\mathrm{m}_{\mathrm{A}j}\cdot S^\mathrm{m}_{\mathrm{B}k}\cdot t_{CC}.
\end{align}
To take into account different detector jitters, one arrives at different values of $t_{\Delta jk}$, which require an adaptation of the coincidence window loss of Eq. \eqref{eq:etatcc}:
\begin{align}
   \eta^{t_{CC}}_{jk}=\textrm{erf}\bigg[\sqrt{\textrm{ln}(2)}\cdot\frac{t_{CC}}{t_{\Delta jk}}\bigg].
\end{align}
In this case, Eq. \eqref{eq:CCm} becomes
\begin{align}
    CC^\mathrm{m}=\sum^{2}_{j,k=1}\bigg[\eta^{t_{CC}}_{jk}\cdot CC^\mathrm{t}_{jk}+CC^\mathrm{acc}_{jk}\bigg].
\end{align}
and similarly, Eq. \eqref{eq:CCerr} can be written as
\begin{align}
    CC^\mathrm{err}=\sum^{2}_{j\neq k} \bigg[ \eta^{t_{CC}}_{jk}\cdot CC^\mathrm{t}_{jk}\cdot  e^\mathrm{pol}+CC^\mathrm{acc}_{jk}\bigg].
\end{align}
Here we assume a correlated Bell state ($\phi^{+/-}$) in the respective basis. For anticorrelated ones ($\psi^{+/-}$), the indices to be summed over have to be replaced by $j=k$.
\subsection{Key-rate-formula adjustments\label{sec:rate}}

Following from above considerations, in a realistic experiment, one might additionally expect that one of the polarization measurement settings used in the BBM92 protocol is more prone to errors than the other one. 
Let us assume that this is due to different optical errors $e^\mathrm{pol}$ which can depend on the measurement basis.
As an example, the $HV$ basis often shows higher fidelity than the superposition bases as a result of the source design, which relies on polarizing beam splitters defining $H$ and $V$ with high extinction ($1:1000$ or better).
Because of this, we obtain two values of QBER [see Eq. \eqref{eq:QBER}], one for each measurement setting. 
Let us denote these with $E_{HV}$ and $E_{DA}$. 
If coincidences obtained in the $HV$ basis are used to derive the key, then in Eq. \eqref{eq:seckeyBBM92} we can set $E_{\textrm{bit}} = E_{HV}$ and $E_{\textrm{ph}} = E_{DA}$. 
Similarly, for a key derived from coincidences in the DA basis we set $E_{\textrm{bit}} = E_{DA}$ and $E_{\textrm{ph}} = E_{HV}$.
If both Alice and Bob choose the $HV$ setting with probability $p$ and the $DA$ setting with probability $(1-p)$, they would obtain two key rates, each in one basis:
\begin{align}
    R^s_{HV}\!&= p^2 CC^m \Big[\!1\!-\!\mathrm{H}_2(E_{DA})\!-\!f(E_{HV})\mathrm{H}_2(E_{HV})\!\Big],\\
    R^s_{DA}\!&= (1\!-\!p)^2 CC^m \Big[\!1\!-\!\mathrm{H}_2(E_{HV})\!-\!f(E_{DA})\mathrm{H}_2(E_{DA})\!\Big].
\end{align}
The total key rate is then the sum of these two key rates, and the total compatible basis choice probability from Eq. \eqref{eq:seckeyBBM92} is $q=p^2+(1-p)^2$.

Another common technique is to use predominantly one of the basis settings and use the other only with very low probability to obtain the estimate on $E_{\mathrm{ph}}$. This is often referred to as the ``efficient BB84 protocol" \cite{Lo2005}. In the asymptotic setting, one can therefore assume that the probability $p$ to choose the $HV$ basis approaches unity, and the final key rate is:
\begin{align}
     R^s_{\mathrm{efficient}} &= CC^m \Big[1-\mathrm{H}_2(E_{DA})-f(E_{HV})\mathrm{H}_2(E_{HV})\Big].
\end{align}

Additionally, in some works the authors assume that in the asymptotic setting the block length is also approaching infinity and therefore $f(E_{\mathrm{bit}})$ approaches unity \cite{PhysRevLett.90.057902,Koashi_2009}. 
Last but not least, even in case of different error rates, one can in practice use the average error $E = (E_{HV}+E_{DA})/2$ with Eq. \eqref{eq:seckey} to obtain a lower bound on the secret key rate \cite{Fung2010, Joshi2019}, since
\begin{align}
    2\cdot \mathrm{H}_2\bigg(\frac{E_1+E_2}{2}\bigg)\geq \mathrm{H}_2(E_1)+\mathrm{H}_2(E_2)\hspace{0.3cm} \forall\, E_i\in\left[0,0.5\right].
\end{align}

\bibliography{CWQKDmodel}

\providecommand{\noopsort}[1]{}\providecommand{\singleletter}[1]{#1}%
\begin{thebibliography}{33}%
\makeatletter
\providecommand \@ifxundefined [1]{%
 \@ifx{#1\undefined}
}%
\providecommand \@ifnum [1]{%
 \ifnum #1\expandafter \@firstoftwo
 \else \expandafter \@secondoftwo
 \fi
}%
\providecommand \@ifx [1]{%
 \ifx #1\expandafter \@firstoftwo
 \else \expandafter \@secondoftwo
 \fi
}%
\providecommand \natexlab [1]{#1}%
\providecommand \enquote  [1]{``#1''}%
\providecommand \bibnamefont  [1]{#1}%
\providecommand \bibfnamefont [1]{#1}%
\providecommand \citenamefont [1]{#1}%
\providecommand \href@noop [0]{\@secondoftwo}%
\providecommand \href [0]{\begingroup \@sanitize@url \@href}%
\providecommand \@href[1]{\@@startlink{#1}\@@href}%
\providecommand \@@href[1]{\endgroup#1\@@endlink}%
\providecommand \@sanitize@url [0]{\catcode `\\12\catcode `\$12\catcode
  `\&12\catcode `\#12\catcode `\^12\catcode `\_12\catcode `\%12\relax}%
\providecommand \@@startlink[1]{}%
\providecommand \@@endlink[0]{}%
\providecommand \url  [0]{\begingroup\@sanitize@url \@url }%
\providecommand \@url [1]{\endgroup\@href {#1}{\urlprefix }}%
\providecommand \urlprefix  [0]{URL }%
\providecommand \Eprint [0]{\href }%
\providecommand \doibase [0]{http://dx.doi.org/}%
\providecommand \selectlanguage [0]{\@gobble}%
\providecommand \bibinfo  [0]{\@secondoftwo}%
\providecommand \bibfield  [0]{\@secondoftwo}%
\providecommand \translation [1]{[#1]}%
\providecommand \BibitemOpen [0]{}%
\providecommand \bibitemStop [0]{}%
\providecommand \bibitemNoStop [0]{.\EOS\space}%
\providecommand \EOS [0]{\spacefactor3000\relax}%
\providecommand \BibitemShut  [1]{\csname bibitem#1\endcsname}%
\let\auto@bib@innerbib\@empty
\bibitem [{\citenamefont {Gisin}\ \emph {et~al.}(2002)\citenamefont {Gisin},
  \citenamefont {Ribordy}, \citenamefont {Tittel},\ and\ \citenamefont
  {Zbinden}}]{Gisin2002}%
  \BibitemOpen
  \bibfield  {author} {\bibinfo {author} {\bibfnamefont {N.}~\bibnamefont
  {Gisin}}, \bibinfo {author} {\bibfnamefont {G.}~\bibnamefont {Ribordy}},
  \bibinfo {author} {\bibfnamefont {W.}~\bibnamefont {Tittel}}, \ and\ \bibinfo
  {author} {\bibfnamefont {H.}~\bibnamefont {Zbinden}},\ }\href@noop {}
  {\bibfield  {journal} {\bibinfo  {journal} {Review of Modern Physics}\
  }\textbf {\bibinfo {volume} {74}},\ \bibinfo {pages} {145} (\bibinfo {year}
  {2002})}\BibitemShut {NoStop}%
\bibitem [{\citenamefont {Xu}\ \emph {et~al.}(2020)\citenamefont {Xu},
  \citenamefont {Ma}, \citenamefont {Zhang}, \citenamefont {Lo},\ and\
  \citenamefont {Pan}}]{Xu2020}%
  \BibitemOpen
  \bibfield  {author} {\bibinfo {author} {\bibfnamefont {F.}~\bibnamefont
  {Xu}}, \bibinfo {author} {\bibfnamefont {X.}~\bibnamefont {Ma}}, \bibinfo
  {author} {\bibfnamefont {Q.}~\bibnamefont {Zhang}}, \bibinfo {author}
  {\bibfnamefont {H.-K.}\ \bibnamefont {Lo}}, \ and\ \bibinfo {author}
  {\bibfnamefont {J.-W.}\ \bibnamefont {Pan}},\ }\href {\doibase
  10.1103/RevModPhys.92.025002} {\bibfield  {journal} {\bibinfo  {journal}
  {Rev. Mod. Phys.}\ }\textbf {\bibinfo {volume} {92}},\ \bibinfo {pages}
  {025002} (\bibinfo {year} {2020})}\BibitemShut {NoStop}%
\bibitem [{\citenamefont {Bennett~Ch}\ and\ \citenamefont
  {Brassard}(1984)}]{Bennett2014}%
  \BibitemOpen
  \bibfield  {author} {\bibinfo {author} {\bibfnamefont {H.}~\bibnamefont
  {Bennett~Ch}}\ and\ \bibinfo {author} {\bibfnamefont {G.}~\bibnamefont
  {Brassard}},\ }in\ \href@noop {} {\emph {\bibinfo {booktitle} {Conf. on
  Computers, Systems and Signal Processing (Bangalore, India, Dec. 1984)}}}\
  (\bibinfo {year} {1984})\ pp.\ \bibinfo {pages} {175--9}\BibitemShut
  {NoStop}%
\bibitem [{\citenamefont {Ursin}\ \emph {et~al.}(2006)\citenamefont {Ursin},
  \citenamefont {Tiefenbacher}, \citenamefont {Schmitt-Manderbach},
  \citenamefont {Weier}, \citenamefont {Scheidl}, \citenamefont {Lindenthal},
  \citenamefont {Blauensteiner}, \citenamefont {Jennewein}, \citenamefont
  {Perdigues}, \citenamefont {Trojek} \emph {et~al.}}]{Ursin2006}%
  \BibitemOpen
  \bibfield  {author} {\bibinfo {author} {\bibfnamefont {R.}~\bibnamefont
  {Ursin}}, \bibinfo {author} {\bibfnamefont {F.}~\bibnamefont {Tiefenbacher}},
  \bibinfo {author} {\bibfnamefont {T.}~\bibnamefont {Schmitt-Manderbach}},
  \bibinfo {author} {\bibfnamefont {H.}~\bibnamefont {Weier}}, \bibinfo
  {author} {\bibfnamefont {T.}~\bibnamefont {Scheidl}}, \bibinfo {author}
  {\bibfnamefont {M.}~\bibnamefont {Lindenthal}}, \bibinfo {author}
  {\bibfnamefont {B.}~\bibnamefont {Blauensteiner}}, \bibinfo {author}
  {\bibfnamefont {T.}~\bibnamefont {Jennewein}}, \bibinfo {author}
  {\bibfnamefont {J.}~\bibnamefont {Perdigues}}, \bibinfo {author}
  {\bibfnamefont {P.}~\bibnamefont {Trojek}},  \emph {et~al.},\ }\href@noop {}
  {\bibfield  {journal} {\bibinfo  {journal} {arXiv preprint quant-ph/0607182}\
  } (\bibinfo {year} {2006})}\BibitemShut {NoStop}%
\bibitem [{\citenamefont {Yin}\ \emph {et~al.}(2012)\citenamefont {Yin},
  \citenamefont {Ren}, \citenamefont {Lu}, \citenamefont {Cao}, \citenamefont
  {Yong}, \citenamefont {Wu}, \citenamefont {Liu}, \citenamefont {Liao},
  \citenamefont {Zhou}, \citenamefont {Jiang}, \citenamefont {Cai},
  \citenamefont {Xu}, \citenamefont {Pan}, \citenamefont {Jia}, \citenamefont
  {Huang}, \citenamefont {Yin}, \citenamefont {Wang}, \citenamefont {Chen},
  \citenamefont {Peng},\ and\ \citenamefont {Pan}}]{Yin2012}%
  \BibitemOpen
  \bibfield  {author} {\bibinfo {author} {\bibfnamefont {J.}~\bibnamefont
  {Yin}}, \bibinfo {author} {\bibfnamefont {J.-G.}\ \bibnamefont {Ren}},
  \bibinfo {author} {\bibfnamefont {H.}~\bibnamefont {Lu}}, \bibinfo {author}
  {\bibfnamefont {Y.}~\bibnamefont {Cao}}, \bibinfo {author} {\bibfnamefont
  {H.-L.}\ \bibnamefont {Yong}}, \bibinfo {author} {\bibfnamefont {Y.-P.}\
  \bibnamefont {Wu}}, \bibinfo {author} {\bibfnamefont {C.}~\bibnamefont
  {Liu}}, \bibinfo {author} {\bibfnamefont {S.-K.}\ \bibnamefont {Liao}},
  \bibinfo {author} {\bibfnamefont {F.}~\bibnamefont {Zhou}}, \bibinfo {author}
  {\bibfnamefont {Y.}~\bibnamefont {Jiang}}, \bibinfo {author} {\bibfnamefont
  {X.-D.}\ \bibnamefont {Cai}}, \bibinfo {author} {\bibfnamefont
  {P.}~\bibnamefont {Xu}}, \bibinfo {author} {\bibfnamefont {G.-S.}\
  \bibnamefont {Pan}}, \bibinfo {author} {\bibfnamefont {J.-J.}\ \bibnamefont
  {Jia}}, \bibinfo {author} {\bibfnamefont {Y.-M.}\ \bibnamefont {Huang}},
  \bibinfo {author} {\bibfnamefont {H.}~\bibnamefont {Yin}}, \bibinfo {author}
  {\bibfnamefont {J.-Y.}\ \bibnamefont {Wang}}, \bibinfo {author}
  {\bibfnamefont {Y.-A.}\ \bibnamefont {Chen}}, \bibinfo {author}
  {\bibfnamefont {C.-Z.}\ \bibnamefont {Peng}}, \ and\ \bibinfo {author}
  {\bibfnamefont {J.-W.}\ \bibnamefont {Pan}},\ }\href {\doibase
  10.1038/nature11332} {\bibfield  {journal} {\bibinfo  {journal} {Nature}\
  }\textbf {\bibinfo {volume} {488}},\ \bibinfo {pages} {185} (\bibinfo {year}
  {2012})}\BibitemShut {NoStop}%
\bibitem [{\citenamefont {Ecker}\ \emph {et~al.}(2021)\citenamefont {Ecker},
  \citenamefont {Liu}, \citenamefont {Handsteiner}, \citenamefont {Fink},
  \citenamefont {Rauch}, \citenamefont {Steinlechner}, \citenamefont {Scheidl},
  \citenamefont {Zeilinger},\ and\ \citenamefont {Ursin}}]{Ecker2021}%
  \BibitemOpen
  \bibfield  {author} {\bibinfo {author} {\bibfnamefont {S.}~\bibnamefont
  {Ecker}}, \bibinfo {author} {\bibfnamefont {B.}~\bibnamefont {Liu}}, \bibinfo
  {author} {\bibfnamefont {J.}~\bibnamefont {Handsteiner}}, \bibinfo {author}
  {\bibfnamefont {M.}~\bibnamefont {Fink}}, \bibinfo {author} {\bibfnamefont
  {D.}~\bibnamefont {Rauch}}, \bibinfo {author} {\bibfnamefont
  {F.}~\bibnamefont {Steinlechner}}, \bibinfo {author} {\bibfnamefont
  {T.}~\bibnamefont {Scheidl}}, \bibinfo {author} {\bibfnamefont
  {A.}~\bibnamefont {Zeilinger}}, \ and\ \bibinfo {author} {\bibfnamefont
  {R.}~\bibnamefont {Ursin}},\ }\href {\doibase 10.1038/s41534-020-00335-5}
  {\bibfield  {journal} {\bibinfo  {journal} {npj Quantum Information}\
  }\textbf {\bibinfo {volume} {7}},\ \bibinfo {pages} {5} (\bibinfo {year}
  {2021})}\BibitemShut {NoStop}%
\bibitem [{\citenamefont {Yin}\ \emph {et~al.}(2020)\citenamefont {Yin},
  \citenamefont {Li}, \citenamefont {Liao}, \citenamefont {Yang}, \citenamefont
  {Cao}, \citenamefont {Zhang}, \citenamefont {Ren}, \citenamefont {Cai},
  \citenamefont {Liu}, \citenamefont {Li}, \citenamefont {Shu}, \citenamefont
  {Huang}, \citenamefont {Deng}, \citenamefont {Li}, \citenamefont {Zhang},
  \citenamefont {Liu}, \citenamefont {Chen}, \citenamefont {Lu}, \citenamefont
  {Wang}, \citenamefont {Xu}, \citenamefont {Wang}, \citenamefont {Peng},
  \citenamefont {Ekert},\ and\ \citenamefont {Pan}}]{Yin2020}%
  \BibitemOpen
  \bibfield  {author} {\bibinfo {author} {\bibfnamefont {J.}~\bibnamefont
  {Yin}}, \bibinfo {author} {\bibfnamefont {Y.-H.}\ \bibnamefont {Li}},
  \bibinfo {author} {\bibfnamefont {S.-K.}\ \bibnamefont {Liao}}, \bibinfo
  {author} {\bibfnamefont {M.}~\bibnamefont {Yang}}, \bibinfo {author}
  {\bibfnamefont {Y.}~\bibnamefont {Cao}}, \bibinfo {author} {\bibfnamefont
  {L.}~\bibnamefont {Zhang}}, \bibinfo {author} {\bibfnamefont {J.-G.}\
  \bibnamefont {Ren}}, \bibinfo {author} {\bibfnamefont {W.-Q.}\ \bibnamefont
  {Cai}}, \bibinfo {author} {\bibfnamefont {W.-Y.}\ \bibnamefont {Liu}},
  \bibinfo {author} {\bibfnamefont {S.-L.}\ \bibnamefont {Li}}, \bibinfo
  {author} {\bibfnamefont {R.}~\bibnamefont {Shu}}, \bibinfo {author}
  {\bibfnamefont {Y.-M.}\ \bibnamefont {Huang}}, \bibinfo {author}
  {\bibfnamefont {L.}~\bibnamefont {Deng}}, \bibinfo {author} {\bibfnamefont
  {L.}~\bibnamefont {Li}}, \bibinfo {author} {\bibfnamefont {Q.}~\bibnamefont
  {Zhang}}, \bibinfo {author} {\bibfnamefont {N.-L.}\ \bibnamefont {Liu}},
  \bibinfo {author} {\bibfnamefont {Y.-A.}\ \bibnamefont {Chen}}, \bibinfo
  {author} {\bibfnamefont {C.-Y.}\ \bibnamefont {Lu}}, \bibinfo {author}
  {\bibfnamefont {X.-B.}\ \bibnamefont {Wang}}, \bibinfo {author}
  {\bibfnamefont {F.}~\bibnamefont {Xu}}, \bibinfo {author} {\bibfnamefont
  {J.-Y.}\ \bibnamefont {Wang}}, \bibinfo {author} {\bibfnamefont {C.-Z.}\
  \bibnamefont {Peng}}, \bibinfo {author} {\bibfnamefont {A.~K.}\ \bibnamefont
  {Ekert}}, \ and\ \bibinfo {author} {\bibfnamefont {J.-W.}\ \bibnamefont
  {Pan}},\ }\href {\doibase 10.1038/s41586-020-2401-y} {\bibfield  {journal}
  {\bibinfo  {journal} {Nature}\ }\textbf {\bibinfo {volume} {582}},\ \bibinfo
  {pages} {501} (\bibinfo {year} {2020})}\BibitemShut {NoStop}%
\bibitem [{\citenamefont {Wengerowsky}\ \emph {et~al.}(2019)\citenamefont
  {Wengerowsky}, \citenamefont {Joshi}, \citenamefont {Steinlechner},
  \citenamefont {Zichi}, \citenamefont {Dobrovolskiy}, \citenamefont {van~der
  Molen}, \citenamefont {Los}, \citenamefont {Zwiller}, \citenamefont
  {Versteegh}, \citenamefont {Mura}, \citenamefont {Calonico}, \citenamefont
  {Inguscio}, \citenamefont {H{\"u}bel}, \citenamefont {Bo}, \citenamefont
  {Scheidl}, \citenamefont {Zeilinger}, \citenamefont {Xuereb},\ and\
  \citenamefont {Ursin}}]{Wengerowsky2019}%
  \BibitemOpen
  \bibfield  {author} {\bibinfo {author} {\bibfnamefont {S.}~\bibnamefont
  {Wengerowsky}}, \bibinfo {author} {\bibfnamefont {S.~K.}\ \bibnamefont
  {Joshi}}, \bibinfo {author} {\bibfnamefont {F.}~\bibnamefont {Steinlechner}},
  \bibinfo {author} {\bibfnamefont {J.~R.}\ \bibnamefont {Zichi}}, \bibinfo
  {author} {\bibfnamefont {S.~M.}\ \bibnamefont {Dobrovolskiy}}, \bibinfo
  {author} {\bibfnamefont {R.}~\bibnamefont {van~der Molen}}, \bibinfo {author}
  {\bibfnamefont {J.~W.~N.}\ \bibnamefont {Los}}, \bibinfo {author}
  {\bibfnamefont {V.}~\bibnamefont {Zwiller}}, \bibinfo {author} {\bibfnamefont
  {M.~A.~M.}\ \bibnamefont {Versteegh}}, \bibinfo {author} {\bibfnamefont
  {A.}~\bibnamefont {Mura}}, \bibinfo {author} {\bibfnamefont {D.}~\bibnamefont
  {Calonico}}, \bibinfo {author} {\bibfnamefont {M.}~\bibnamefont {Inguscio}},
  \bibinfo {author} {\bibfnamefont {H.}~\bibnamefont {H{\"u}bel}}, \bibinfo
  {author} {\bibfnamefont {L.}~\bibnamefont {Bo}}, \bibinfo {author}
  {\bibfnamefont {T.}~\bibnamefont {Scheidl}}, \bibinfo {author} {\bibfnamefont
  {A.}~\bibnamefont {Zeilinger}}, \bibinfo {author} {\bibfnamefont
  {A.}~\bibnamefont {Xuereb}}, \ and\ \bibinfo {author} {\bibfnamefont
  {R.}~\bibnamefont {Ursin}},\ }\href {\doibase 10.1073/pnas.1818752116}
  {\bibfield  {journal} {\bibinfo  {journal} {Proceedings of the National
  Academy of Sciences}\ }\textbf {\bibinfo {volume} {116}},\ \bibinfo {pages}
  {6684} (\bibinfo {year} {2019})},\ \Eprint
  {http://arxiv.org/abs/https://www.pnas.org/content/116/14/6684.full.pdf}
  {https://www.pnas.org/content/116/14/6684.full.pdf} \BibitemShut {NoStop}%
\bibitem [{\citenamefont {Wengerowsky}\ \emph {et~al.}(2018)\citenamefont
  {Wengerowsky}, \citenamefont {Joshi}, \citenamefont {Steinlechner},
  \citenamefont {H{\"u}bel},\ and\ \citenamefont {Ursin}}]{Wengerowsky2018a}%
  \BibitemOpen
  \bibfield  {author} {\bibinfo {author} {\bibfnamefont {S.}~\bibnamefont
  {Wengerowsky}}, \bibinfo {author} {\bibfnamefont {S.~K.}\ \bibnamefont
  {Joshi}}, \bibinfo {author} {\bibfnamefont {F.}~\bibnamefont {Steinlechner}},
  \bibinfo {author} {\bibfnamefont {H.}~\bibnamefont {H{\"u}bel}}, \ and\
  \bibinfo {author} {\bibfnamefont {R.}~\bibnamefont {Ursin}},\ }\href@noop {}
  {\bibfield  {journal} {\bibinfo  {journal} {Nature}\ }\textbf {\bibinfo
  {volume} {564}},\ \bibinfo {pages} {225} (\bibinfo {year}
  {2018})}\BibitemShut {NoStop}%
\bibitem [{\citenamefont {Joshi}\ \emph {et~al.}(2019)\citenamefont {Joshi},
  \citenamefont {Aktas}, \citenamefont {Wengerowsky}, \citenamefont
  {Lončarić}, \citenamefont {Neumann}, \citenamefont {Liu}, \citenamefont
  {Scheidl}, \citenamefont {Željko Samec}, \citenamefont {Kling},
  \citenamefont {Qiu}, \citenamefont {Stippčević}, \citenamefont {Rarity},\
  and\ \citenamefont {Ursin}}]{Joshi2019}%
  \BibitemOpen
  \bibfield  {author} {\bibinfo {author} {\bibfnamefont {S.~K.}\ \bibnamefont
  {Joshi}}, \bibinfo {author} {\bibfnamefont {D.}~\bibnamefont {Aktas}},
  \bibinfo {author} {\bibfnamefont {S.}~\bibnamefont {Wengerowsky}}, \bibinfo
  {author} {\bibfnamefont {M.}~\bibnamefont {Lončarić}}, \bibinfo {author}
  {\bibfnamefont {S.~P.}\ \bibnamefont {Neumann}}, \bibinfo {author}
  {\bibfnamefont {B.}~\bibnamefont {Liu}}, \bibinfo {author} {\bibfnamefont
  {T.}~\bibnamefont {Scheidl}}, \bibinfo {author} {\bibnamefont {Željko
  Samec}}, \bibinfo {author} {\bibfnamefont {L.}~\bibnamefont {Kling}},
  \bibinfo {author} {\bibfnamefont {A.}~\bibnamefont {Qiu}}, \bibinfo {author}
  {\bibfnamefont {M.}~\bibnamefont {Stippčević}}, \bibinfo {author}
  {\bibfnamefont {J.~G.}\ \bibnamefont {Rarity}}, \ and\ \bibinfo {author}
  {\bibfnamefont {R.}~\bibnamefont {Ursin}},\ }\href@noop {} {\enquote
  {\bibinfo {title} {A trusted-node-free eight-user metropolitan quantum
  communication network},}\ } (\bibinfo {year} {2019}),\ \Eprint
  {http://arxiv.org/abs/1907.08229} {arXiv:1907.08229 [quant-ph]} \BibitemShut
  {NoStop}%
\bibitem [{\citenamefont {Ekert}(1991)}]{Ekert1991}%
  \BibitemOpen
  \bibfield  {author} {\bibinfo {author} {\bibfnamefont {A.~K.}\ \bibnamefont
  {Ekert}},\ }\href@noop {} {\bibfield  {journal} {\bibinfo  {journal}
  {Physical review letters}\ }\textbf {\bibinfo {volume} {67}},\ \bibinfo
  {pages} {661} (\bibinfo {year} {1991})}\BibitemShut {NoStop}%
\bibitem [{\citenamefont {Bennett}\ \emph {et~al.}(1992)\citenamefont
  {Bennett}, \citenamefont {Brassard},\ and\ \citenamefont
  {Mermin}}]{Bennett1992}%
  \BibitemOpen
  \bibfield  {author} {\bibinfo {author} {\bibfnamefont {C.~H.}\ \bibnamefont
  {Bennett}}, \bibinfo {author} {\bibfnamefont {G.}~\bibnamefont {Brassard}}, \
  and\ \bibinfo {author} {\bibfnamefont {N.~D.}\ \bibnamefont {Mermin}},\
  }\href@noop {} {\bibfield  {journal} {\bibinfo  {journal} {Physical Review
  Letters}\ }\textbf {\bibinfo {volume} {68}},\ \bibinfo {pages} {557}
  (\bibinfo {year} {1992})}\BibitemShut {NoStop}%
\bibitem [{\citenamefont {Chen}\ \emph {et~al.}(2021)\citenamefont {Chen},
  \citenamefont {Zhang}, \citenamefont {Liu}, \citenamefont {Jiang},
  \citenamefont {Zhang}, \citenamefont {Han}, \citenamefont {Ma}, \citenamefont
  {Hu}, \citenamefont {Li}, \citenamefont {Liu}, \citenamefont {Zhou},
  \citenamefont {Jiang}, \citenamefont {Chen}, \citenamefont {Li},
  \citenamefont {You}, \citenamefont {Wang}, \citenamefont {Wang},
  \citenamefont {Zhang},\ and\ \citenamefont {Pan}}]{Chen2021arx}%
  \BibitemOpen
  \bibfield  {author} {\bibinfo {author} {\bibfnamefont {J.-P.}\ \bibnamefont
  {Chen}}, \bibinfo {author} {\bibfnamefont {C.}~\bibnamefont {Zhang}},
  \bibinfo {author} {\bibfnamefont {Y.}~\bibnamefont {Liu}}, \bibinfo {author}
  {\bibfnamefont {C.}~\bibnamefont {Jiang}}, \bibinfo {author} {\bibfnamefont
  {W.-J.}\ \bibnamefont {Zhang}}, \bibinfo {author} {\bibfnamefont {Z.-Y.}\
  \bibnamefont {Han}}, \bibinfo {author} {\bibfnamefont {S.-Z.}\ \bibnamefont
  {Ma}}, \bibinfo {author} {\bibfnamefont {X.-L.}\ \bibnamefont {Hu}}, \bibinfo
  {author} {\bibfnamefont {Y.-H.}\ \bibnamefont {Li}}, \bibinfo {author}
  {\bibfnamefont {H.}~\bibnamefont {Liu}}, \bibinfo {author} {\bibfnamefont
  {F.}~\bibnamefont {Zhou}}, \bibinfo {author} {\bibfnamefont {H.-F.}\
  \bibnamefont {Jiang}}, \bibinfo {author} {\bibfnamefont {T.-Y.}\ \bibnamefont
  {Chen}}, \bibinfo {author} {\bibfnamefont {H.}~\bibnamefont {Li}}, \bibinfo
  {author} {\bibfnamefont {L.-X.}\ \bibnamefont {You}}, \bibinfo {author}
  {\bibfnamefont {Z.}~\bibnamefont {Wang}}, \bibinfo {author} {\bibfnamefont
  {X.-B.}\ \bibnamefont {Wang}}, \bibinfo {author} {\bibfnamefont
  {Q.}~\bibnamefont {Zhang}}, \ and\ \bibinfo {author} {\bibfnamefont {J.-W.}\
  \bibnamefont {Pan}},\ }\href@noop {} {\enquote {\bibinfo {title} {Twin-field
  quantum key distribution over 511 km optical fiber linking two distant
  metropolitans},}\ } (\bibinfo {year} {2021}),\ \Eprint
  {http://arxiv.org/abs/2102.00433} {arXiv:2102.00433 [quant-ph]} \BibitemShut
  {NoStop}%
\bibitem [{\citenamefont {Lo}\ \emph {et~al.}(2005{\natexlab{a}})\citenamefont
  {Lo}, \citenamefont {Ma},\ and\ \citenamefont {Chen}}]{Lo05a}%
  \BibitemOpen
  \bibfield  {author} {\bibinfo {author} {\bibfnamefont {H.-K.}\ \bibnamefont
  {Lo}}, \bibinfo {author} {\bibfnamefont {X.}~\bibnamefont {Ma}}, \ and\
  \bibinfo {author} {\bibfnamefont {K.}~\bibnamefont {Chen}},\ }\href {\doibase
  10.1103/PhysRevLett.94.230504} {\bibfield  {journal} {\bibinfo  {journal}
  {Phys. Rev. Lett.}\ }\textbf {\bibinfo {volume} {94}},\ \bibinfo {pages}
  {230504} (\bibinfo {year} {2005}{\natexlab{a}})}\BibitemShut {NoStop}%
\bibitem [{\citenamefont {Ac\'{\i}n}\ \emph {et~al.}(2007)\citenamefont
  {Ac\'{\i}n}, \citenamefont {Brunner}, \citenamefont {Gisin}, \citenamefont
  {Massar}, \citenamefont {Pironio},\ and\ \citenamefont {Scarani}}]{Acin2007}%
  \BibitemOpen
  \bibfield  {author} {\bibinfo {author} {\bibfnamefont {A.}~\bibnamefont
  {Ac\'{\i}n}}, \bibinfo {author} {\bibfnamefont {N.}~\bibnamefont {Brunner}},
  \bibinfo {author} {\bibfnamefont {N.}~\bibnamefont {Gisin}}, \bibinfo
  {author} {\bibfnamefont {S.}~\bibnamefont {Massar}}, \bibinfo {author}
  {\bibfnamefont {S.}~\bibnamefont {Pironio}}, \ and\ \bibinfo {author}
  {\bibfnamefont {V.}~\bibnamefont {Scarani}},\ }\href {\doibase
  10.1103/PhysRevLett.98.230501} {\bibfield  {journal} {\bibinfo  {journal}
  {Phys. Rev. Lett.}\ }\textbf {\bibinfo {volume} {98}},\ \bibinfo {pages}
  {230501} (\bibinfo {year} {2007})}\BibitemShut {NoStop}%
\bibitem [{\citenamefont {Ma}\ \emph {et~al.}(2007)\citenamefont {Ma},
  \citenamefont {Fung},\ and\ \citenamefont {Lo}}]{Ma2007}%
  \BibitemOpen
  \bibfield  {author} {\bibinfo {author} {\bibfnamefont {X.}~\bibnamefont
  {Ma}}, \bibinfo {author} {\bibfnamefont {C.~H.~F.}\ \bibnamefont {Fung}}, \
  and\ \bibinfo {author} {\bibfnamefont {H.~K.}\ \bibnamefont {Lo}},\
  }\href@noop {} {\bibfield  {journal} {\bibinfo  {journal} {Physical Review
  A}\ }\textbf {\bibinfo {volume} {76}},\ \bibinfo {pages} {012307} (\bibinfo
  {year} {2007})}\BibitemShut {NoStop}%
\bibitem [{\citenamefont {Neumann}\ \emph {et~al.}(2021)\citenamefont
  {Neumann}, \citenamefont {Ribezzo}, \citenamefont {Bohmann},\ and\
  \citenamefont {Ursin}}]{Neumann2021}%
  \BibitemOpen
  \bibfield  {author} {\bibinfo {author} {\bibfnamefont {S.~P.}\ \bibnamefont
  {Neumann}}, \bibinfo {author} {\bibfnamefont {D.}~\bibnamefont {Ribezzo}},
  \bibinfo {author} {\bibfnamefont {M.}~\bibnamefont {Bohmann}}, \ and\
  \bibinfo {author} {\bibfnamefont {R.}~\bibnamefont {Ursin}},\ }\href
  {http://iopscience.iop.org/article/10.1088/2058-9565/abe5ee} {\bibfield
  {journal} {\bibinfo  {journal} {Quantum Science and Technology}\ } (\bibinfo
  {year} {2021})}\BibitemShut {NoStop}%
\bibitem [{\citenamefont {Takesue}\ and\ \citenamefont
  {Shimizu}(2010)}]{Takesue2010}%
  \BibitemOpen
  \bibfield  {author} {\bibinfo {author} {\bibfnamefont {H.}~\bibnamefont
  {Takesue}}\ and\ \bibinfo {author} {\bibfnamefont {K.}~\bibnamefont
  {Shimizu}},\ }\href@noop {} {\bibfield  {journal} {\bibinfo  {journal}
  {Optics Communications}\ }\textbf {\bibinfo {volume} {283}},\ \bibinfo
  {pages} {276} (\bibinfo {year} {2010})}\BibitemShut {NoStop}%
\bibitem [{\citenamefont {Klyshko}(1980)}]{Klyshko1980}%
  \BibitemOpen
  \bibfield  {author} {\bibinfo {author} {\bibfnamefont {D.~N.}\ \bibnamefont
  {Klyshko}},\ }\href {\doibase 10.1070/qe1980v010n09abeh010660} {\bibfield
  {journal} {\bibinfo  {journal} {Soviet Journal of Quantum Electronics}\
  }\textbf {\bibinfo {volume} {10}},\ \bibinfo {pages} {1112} (\bibinfo {year}
  {1980})}\BibitemShut {NoStop}%
\bibitem [{\citenamefont {Anwar}\ \emph {et~al.}(2021)\citenamefont {Anwar},
  \citenamefont {Perumangatt}, \citenamefont {Steinlechner}, \citenamefont
  {Jennewein},\ and\ \citenamefont {Ling}}]{Anwar2021}%
  \BibitemOpen
  \bibfield  {author} {\bibinfo {author} {\bibfnamefont {A.}~\bibnamefont
  {Anwar}}, \bibinfo {author} {\bibfnamefont {C.}~\bibnamefont {Perumangatt}},
  \bibinfo {author} {\bibfnamefont {F.}~\bibnamefont {Steinlechner}}, \bibinfo
  {author} {\bibfnamefont {T.}~\bibnamefont {Jennewein}}, \ and\ \bibinfo
  {author} {\bibfnamefont {A.}~\bibnamefont {Ling}},\ }\href {\doibase
  10.1063/5.0023103} {\bibfield  {journal} {\bibinfo  {journal} {Review of
  Scientific Instruments}\ }\textbf {\bibinfo {volume} {92}},\ \bibinfo {pages}
  {041101} (\bibinfo {year} {2021})},\ \Eprint
  {http://arxiv.org/abs/https://doi.org/10.1063/5.0023103}
  {https://doi.org/10.1063/5.0023103} \BibitemShut {NoStop}%
\bibitem [{\citenamefont {M.Bohmann}\ \emph {et~al.}(2017)\citenamefont
  {M.Bohmann}, \citenamefont {Kruse}, \citenamefont {Sperling}, \citenamefont
  {Silberhorn},\ and\ \citenamefont {Vogel}}]{Bohmann2017}%
  \BibitemOpen
  \bibfield  {author} {\bibinfo {author} {\bibnamefont {M.Bohmann}}, \bibinfo
  {author} {\bibfnamefont {R.}~\bibnamefont {Kruse}}, \bibinfo {author}
  {\bibfnamefont {J.}~\bibnamefont {Sperling}}, \bibinfo {author}
  {\bibfnamefont {C.}~\bibnamefont {Silberhorn}}, \ and\ \bibinfo {author}
  {\bibfnamefont {W.}~\bibnamefont {Vogel}},\ }\href {\doibase
  10.1103/PhysRevA.95.033806} {\bibfield  {journal} {\bibinfo  {journal} {Phys.
  Rev. A}\ }\textbf {\bibinfo {volume} {95}},\ \bibinfo {pages} {033806}
  (\bibinfo {year} {2017})}\BibitemShut {NoStop}%
\bibitem [{\citenamefont {Bohmann}\ \emph {et~al.}(2019)\citenamefont
  {Bohmann}, \citenamefont {Qi}, \citenamefont {Vogel},\ and\ \citenamefont
  {Chekhova}}]{Bohmann2019}%
  \BibitemOpen
  \bibfield  {author} {\bibinfo {author} {\bibfnamefont {M.}~\bibnamefont
  {Bohmann}}, \bibinfo {author} {\bibfnamefont {L.}~\bibnamefont {Qi}},
  \bibinfo {author} {\bibfnamefont {W.}~\bibnamefont {Vogel}}, \ and\ \bibinfo
  {author} {\bibfnamefont {M.}~\bibnamefont {Chekhova}},\ }\href {\doibase
  10.1103/PhysRevResearch.1.033178} {\bibfield  {journal} {\bibinfo  {journal}
  {Phys. Rev. Research}\ }\textbf {\bibinfo {volume} {1}},\ \bibinfo {pages}
  {033178} (\bibinfo {year} {2019})}\BibitemShut {NoStop}%
\bibitem [{\citenamefont {L\"utkenhaus}(1999)}]{Luetkenhaus1999}%
  \BibitemOpen
  \bibfield  {author} {\bibinfo {author} {\bibfnamefont {N.}~\bibnamefont
  {L\"utkenhaus}},\ }\href {\doibase 10.1103/PhysRevA.59.3301} {\bibfield
  {journal} {\bibinfo  {journal} {Phys. Rev. A}\ }\textbf {\bibinfo {volume}
  {59}},\ \bibinfo {pages} {3301} (\bibinfo {year} {1999})}\BibitemShut
  {NoStop}%
\bibitem [{\citenamefont {Moroder}\ \emph {et~al.}(2010)\citenamefont
  {Moroder}, \citenamefont {G\"uhne}, \citenamefont {Beaudry}, \citenamefont
  {Piani},\ and\ \citenamefont {L\"utkenhaus}}]{Moroder2010}%
  \BibitemOpen
  \bibfield  {author} {\bibinfo {author} {\bibfnamefont {T.}~\bibnamefont
  {Moroder}}, \bibinfo {author} {\bibfnamefont {O.}~\bibnamefont {G\"uhne}},
  \bibinfo {author} {\bibfnamefont {N.}~\bibnamefont {Beaudry}}, \bibinfo
  {author} {\bibfnamefont {M.}~\bibnamefont {Piani}}, \ and\ \bibinfo {author}
  {\bibfnamefont {N.}~\bibnamefont {L\"utkenhaus}},\ }\href {\doibase
  10.1103/PhysRevA.81.052342} {\bibfield  {journal} {\bibinfo  {journal} {Phys.
  Rev. A}\ }\textbf {\bibinfo {volume} {81}},\ \bibinfo {pages} {052342}
  (\bibinfo {year} {2010})}\BibitemShut {NoStop}%
\bibitem [{\citenamefont {Elkouss}\ \emph {et~al.}(2011)\citenamefont
  {Elkouss}, \citenamefont {Mart{\'{\i}}nez{-}Mateo},\ and\ \citenamefont
  {Martin}}]{Elkouss2011}%
  \BibitemOpen
  \bibfield  {author} {\bibinfo {author} {\bibfnamefont {D.}~\bibnamefont
  {Elkouss}}, \bibinfo {author} {\bibfnamefont {J.}~\bibnamefont
  {Mart{\'{\i}}nez{-}Mateo}}, \ and\ \bibinfo {author} {\bibfnamefont
  {V.}~\bibnamefont {Martin}},\ }\href
  {http://www.rintonpress.com/xxqic11/qic-11-34/0226-0238.pdf} {\bibfield
  {journal} {\bibinfo  {journal} {Quantum Inf. Comput.}\ }\textbf {\bibinfo
  {volume} {11}},\ \bibinfo {pages} {226} (\bibinfo {year} {2011})}\BibitemShut
  {NoStop}%
\bibitem [{\citenamefont {Vasylyev}\ \emph {et~al.}(2016)\citenamefont
  {Vasylyev}, \citenamefont {Semenov},\ and\ \citenamefont
  {Vogel}}]{Vasylyev2016}%
  \BibitemOpen
  \bibfield  {author} {\bibinfo {author} {\bibfnamefont {D.}~\bibnamefont
  {Vasylyev}}, \bibinfo {author} {\bibfnamefont {A.~A.}\ \bibnamefont
  {Semenov}}, \ and\ \bibinfo {author} {\bibfnamefont {W.}~\bibnamefont
  {Vogel}},\ }\href {\doibase 10.1103/PhysRevLett.117.090501} {\bibfield
  {journal} {\bibinfo  {journal} {Phys. Rev. Lett.}\ }\textbf {\bibinfo
  {volume} {117}},\ \bibinfo {pages} {090501} (\bibinfo {year}
  {2016})}\BibitemShut {NoStop}%
\bibitem [{\citenamefont {Bohmann}\ \emph {et~al.}(2017)\citenamefont
  {Bohmann}, \citenamefont {Kruse}, \citenamefont {Sperling}, \citenamefont
  {Silberhorn},\ and\ \citenamefont {Vogel}}]{Bohmann2017a}%
  \BibitemOpen
  \bibfield  {author} {\bibinfo {author} {\bibfnamefont {M.}~\bibnamefont
  {Bohmann}}, \bibinfo {author} {\bibfnamefont {R.}~\bibnamefont {Kruse}},
  \bibinfo {author} {\bibfnamefont {J.}~\bibnamefont {Sperling}}, \bibinfo
  {author} {\bibfnamefont {C.}~\bibnamefont {Silberhorn}}, \ and\ \bibinfo
  {author} {\bibfnamefont {W.}~\bibnamefont {Vogel}},\ }\href {\doibase
  10.1103/PhysRevA.95.063801} {\bibfield  {journal} {\bibinfo  {journal} {Phys.
  Rev. A}\ }\textbf {\bibinfo {volume} {95}},\ \bibinfo {pages} {063801}
  (\bibinfo {year} {2017})}\BibitemShut {NoStop}%
\bibitem [{\citenamefont {B{\'e}cares}\ and\ \citenamefont
  {Bl{\'a}zquez}(2012)}]{Becares2012}%
  \BibitemOpen
  \bibfield  {author} {\bibinfo {author} {\bibfnamefont {V.}~\bibnamefont
  {B{\'e}cares}}\ and\ \bibinfo {author} {\bibfnamefont {J.}~\bibnamefont
  {Bl{\'a}zquez}},\ }\href {\doibase 10.1155/2012/240693} {\bibfield  {journal}
  {\bibinfo  {journal} {Science and Technology of Nuclear Installations}\
  }\textbf {\bibinfo {volume} {2012}},\ \bibinfo {pages} {240693} (\bibinfo
  {year} {2012})}\BibitemShut {NoStop}%
\bibitem [{\citenamefont {Zhang}\ \emph {et~al.}(2021)\citenamefont {Zhang},
  \citenamefont {Coles}, \citenamefont {Winick}, \citenamefont {Lin},\ and\
  \citenamefont {L\"utkenhaus}}]{Zhang2021}%
  \BibitemOpen
  \bibfield  {author} {\bibinfo {author} {\bibfnamefont {Y.}~\bibnamefont
  {Zhang}}, \bibinfo {author} {\bibfnamefont {P.~J.}\ \bibnamefont {Coles}},
  \bibinfo {author} {\bibfnamefont {A.}~\bibnamefont {Winick}}, \bibinfo
  {author} {\bibfnamefont {J.}~\bibnamefont {Lin}}, \ and\ \bibinfo {author}
  {\bibfnamefont {N.}~\bibnamefont {L\"utkenhaus}},\ }\href {\doibase
  10.1103/PhysRevResearch.3.013076} {\bibfield  {journal} {\bibinfo  {journal}
  {Phys. Rev. Research}\ }\textbf {\bibinfo {volume} {3}},\ \bibinfo {pages}
  {013076} (\bibinfo {year} {2021})}\BibitemShut {NoStop}%
\bibitem [{\citenamefont {Lo}\ \emph {et~al.}(2005{\natexlab{b}})\citenamefont
  {Lo}, \citenamefont {Chau},\ and\ \citenamefont {Ardehali}}]{Lo2005}%
  \BibitemOpen
  \bibfield  {author} {\bibinfo {author} {\bibfnamefont {H.-K.}\ \bibnamefont
  {Lo}}, \bibinfo {author} {\bibfnamefont {H.~F.}\ \bibnamefont {Chau}}, \ and\
  \bibinfo {author} {\bibfnamefont {M.}~\bibnamefont {Ardehali}},\ }\href
  {\doibase 10.1007/s00145-004-0142-y} {\bibfield  {journal} {\bibinfo
  {journal} {Journal of Cryptology}\ }\textbf {\bibinfo {volume} {18}},\
  \bibinfo {pages} {133} (\bibinfo {year} {2005}{\natexlab{b}})}\BibitemShut
  {NoStop}%
\bibitem [{\citenamefont {Koashi}\ and\ \citenamefont
  {Preskill}(2003)}]{PhysRevLett.90.057902}%
  \BibitemOpen
  \bibfield  {author} {\bibinfo {author} {\bibfnamefont {M.}~\bibnamefont
  {Koashi}}\ and\ \bibinfo {author} {\bibfnamefont {J.}~\bibnamefont
  {Preskill}},\ }\href {\doibase 10.1103/PhysRevLett.90.057902} {\bibfield
  {journal} {\bibinfo  {journal} {Phys. Rev. Lett.}\ }\textbf {\bibinfo
  {volume} {90}},\ \bibinfo {pages} {057902} (\bibinfo {year}
  {2003})}\BibitemShut {NoStop}%
\bibitem [{\citenamefont {Koashi}(2009)}]{Koashi_2009}%
  \BibitemOpen
  \bibfield  {author} {\bibinfo {author} {\bibfnamefont {M.}~\bibnamefont
  {Koashi}},\ }\href {\doibase 10.1088/1367-2630/11/4/045018} {\bibfield
  {journal} {\bibinfo  {journal} {New Journal of Physics}\ }\textbf {\bibinfo
  {volume} {11}},\ \bibinfo {pages} {045018} (\bibinfo {year}
  {2009})}\BibitemShut {NoStop}%
\bibitem [{\citenamefont {Fung}\ \emph {et~al.}(2010)\citenamefont {Fung},
  \citenamefont {Ma},\ and\ \citenamefont {Chau}}]{Fung2010}%
  \BibitemOpen
  \bibfield  {author} {\bibinfo {author} {\bibfnamefont {C.~H.~F.}\
  \bibnamefont {Fung}}, \bibinfo {author} {\bibfnamefont {X.}~\bibnamefont
  {Ma}}, \ and\ \bibinfo {author} {\bibfnamefont {H.~F.}\ \bibnamefont
  {Chau}},\ }\href@noop {} {\bibfield  {journal} {\bibinfo  {journal} {Physical
  Review A}\ }\textbf {\bibinfo {volume} {81}},\ \bibinfo {pages} {012318}
  (\bibinfo {year} {2010})}\BibitemShut {NoStop}%
\end{thebibliography}%

\end{document}